# CAR T Cells from Code to Clinic: Framing Modeling Approaches with Current Translational Research Goals

## Authors

Lucas E Sant'Anna[1,2,3,5,*], Rohita Roy[3,5], Janella C Schwab,[1,4] Julian I Perez,[1,2] and Michaëlle N. Mayalu[3,6]

## Author Affiliations

[1]Department of Bioengineering, Stanford University, Stanford, CA 94305, USA
[2]Department of Genetics, Stanford University School of Medicine, Stanford, CA 94305, USA
[3]Department of Mechanical Engineering, Stanford University, Stanford, CA 94305, USA
[4]Department of Biomedical Data Science, Stanford University School of Medicine, Stanford, CA 94305, USA
[5]These authors contributed equally
[6]Lead contact
*Correspondence: santanna@stanford.edu, mmayalu@stanford.edu

## Acknowledgments

We would like to thank members of the Mayalu and Daniels Lab for supportive and productive discussions. Funding for this work was provided in part by the National Science Foundation Graduate Research Fellowship Program (DGE-2146755; L.E.S. and J.C.S), and the Stanford Graduate Fellowship (J.C.S.).

## Declaration of Interests

The authors declare no competing interests.

**Graphical Abstract**

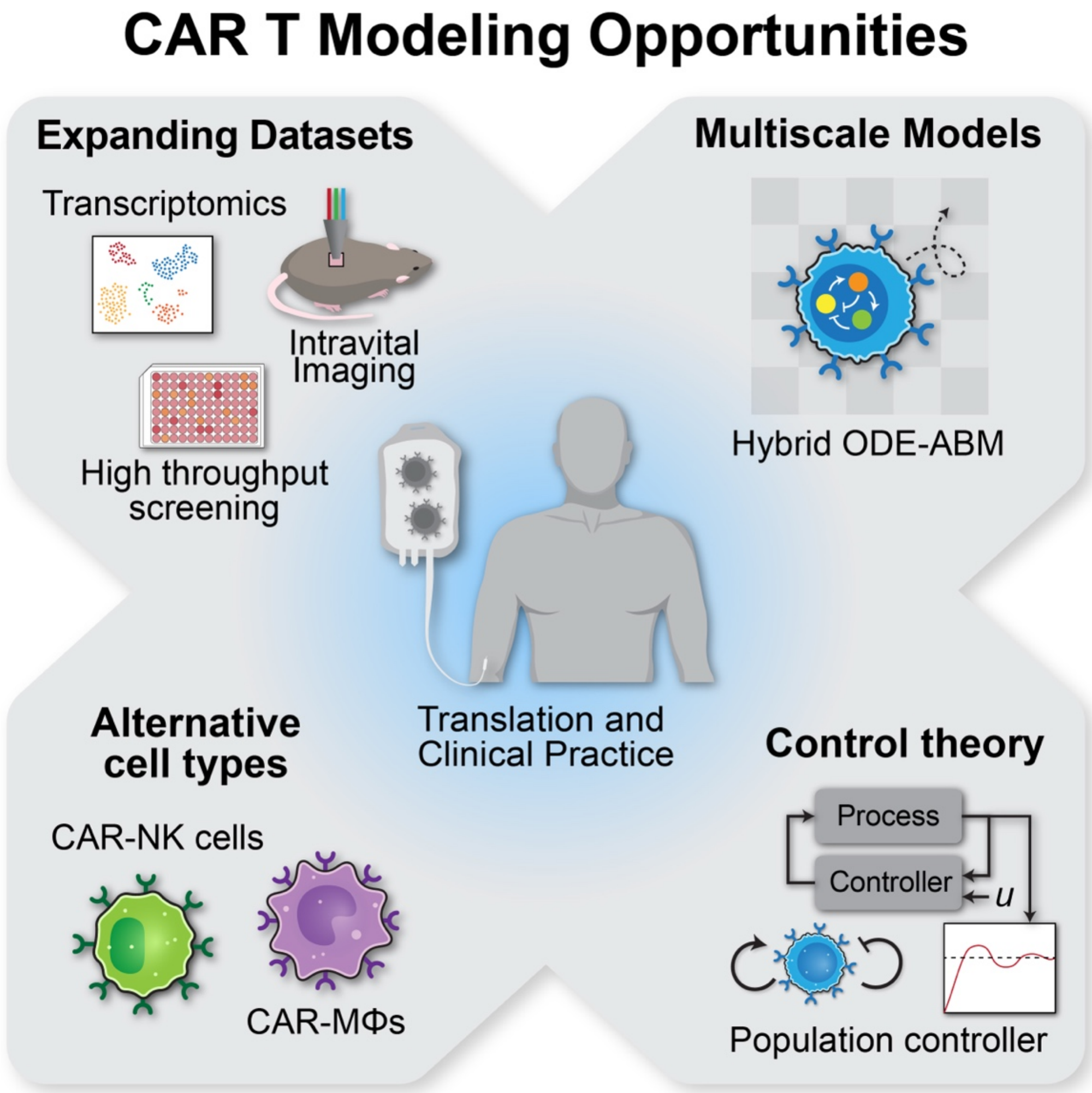

**ABSTRACT**

Chimeric Antigen Receptor (CAR) T cell therapy has transformed immunotherapy for resistant cancers, yet it faces major limitations such as lack of persistence, toxicity, exhaustion, and antigen-negative relapse. Enhancing CAR T cells with genetic circuitry and synthetic receptors offers solutions to some of these problems, but often the theoretical design space is too large to explore experimentally. Mathematical modeling offers a powerful framework for addressing these translational bottlenecks by linking mechanistic understanding to design optimization and clinical application.

This perspective embeds modeling methodologies within the therapeutic problems they aim to solve, framing the discussion around key translational challenges rather than modeling techniques. We critically evaluate the strengths, limitations, and data gaps of current approaches emphasizing how modeling supports the development of safer and more effective therapies. We highlight emerging approaches such as multiscale modeling, control theory, and data-driven methods that leverage high-dimensional datasets to guide predictive design, and we point toward underexplored areas in immune cell therapy including CAR NK and CAR macrophages as future modeling frontiers. We hope that the themes explored in this perspective will encourage readers to refine predictive models, enabling researchers to optimize CAR T cell therapies at the genetic, cellular, microenvironmental, and patient level to enhance their clinical performance.

**Keywords:** CAR T cell therapy, mathematical modeling, multiscale modeling, pharmacokinetics and pharmacodynamics, data-driven design, immune cell engineering, clinical translation, machine learning.

## INTRODUCTION

Cancer remains one of the most difficult diseases to treat, largely due to heterogeneous treatment responses, high relapse rates, and poor outcomes in patients with relapsed or refractory disease.[1] Among recent breakthroughs in oncology, Chimeric Antigen Receptor (CAR) T cell therapy stands out as a powerful new approach, particularly in B-cell hematologic (blood) cancers **(Box 1)**. This therapy involves engineering a patient's T cells to express a synthetic receptor—comprising of an antigen-binding domain that recognizes cancer targets, co-stimulatory domains to enhance activation, and a TCR signaling domain that triggers killing.[2] Since early studies in the late 1980s, decades of research have led to FDA approval of seven CAR T therapies for B-cell lymphomas, leukemias, and multiple myeloma.[2,3] Clinical success of CAR T cell therapy reflects iterative engineering advances, from first-generation constructs to the sophisticated modern designs that include logic-gated expression of CARs; multiple antigen targets; expression of costimulatory domains, cytokine receptors, and transcription factors; genetic knockouts and many other synthetic biological technologies.[4]

In clinical settings, CAR T therapies have shown remarkable response rates. A review of 38 clinical studies in relapsed patients with B-cell cancers found overall response rates of ~72%, underscoring the potential of CAR T cell therapy as a second- or third-line therapy.[5] However, all approved CAR T therapies remain limited to hematologic cancers (<10% of U.S. cases), while clinical success in solid tumors continues to lag, despite over 1,500 trials worldwide.[6]

## The role of modeling in addressing challenges in CAR T cell therapy

Several biological and clinical challenges currently hinder the broader application of CAR T therapies, including T cell exhaustion (dysfunctional state from chronic stimulation), cellular- and tissue-level barriers like poor tumor infiltration, antigen escape (loss of target antigen expression), and systemic adverse effects including cytokine release syndrome (CRS; dangerous excessive immune activation), neurotoxicity, and other treatment-related toxicities as mentioned in **(Table 1).** The table provides a systematic summary of the current challenges in CAR T cell therapy. Many of these obstacles involve complex, context-dependent interactions between the immune system and engineered T cells, making them difficult to resolve using experimental or clinical testing alone.[7] As "living therapeutics," CAR T products link molecular design decisions to cellular phenotypes and patient-specific responses, creating a vast, costly, and time-intensive design space.[8]

Predictive and generalizable mathematical and computational models are well recognized in other domains. In engineering disciplines such as aerospace engineering, civil engineering, and robotics, high-fidelity modeling plays a central role in system design, analysis, and optimization. Similarly, in pharmacological research outside of cell therapies, model-informed drug development (MIDD) employs mathematical frameworks to characterize pharmacokinetics (PK), pharmacodynamics (PD), and disease progression. The FDA has established formal pathways for the use of MIDD in use regulatory decision-making, underscoring its translational impact.[9] Mathematical modeling offers a scalable, hypothesis-driven way to accelerate CAR T design, refine therapeutic strategies, interpret data, and predict outcomes *in silico*, while integration with experiments helps navigate biological complexity and advance new designs more efficiently and safely to the clinic.[10]

However, despite significant advances in modeling tools available to address and explore limitations of CAR T cell therapy, accurately capturing its biological complexities remains a major hurdle. Accurately reflecting nonlinear, stochastic, and multi-scale biology remains difficult, with challenges including dynamic CAR T states, immune crosstalk, and spatial barriers in solid tumors (**Table 2**).

## Mathematical and Computational Tools in Modeling CAR T Cell Therapy

Before discussing specific recent advances, it is important to note that addressing the aforementioned challenges requires predictive and/or mechanism-based models that we describe below. Researchers are increasingly applying mathematical and computational tools, including differential equations, rule-based logic, stochastic processes, and statistical learning to capture CAR T dynamics across molecular, cellular, and tissue scales. These tools enable modeling approaches for the analysis of key mechanisms of CAR T efficacy, exploration of design alternatives, and prediction of therapeutic outcomes.

### *Differential Equation Tools*

Ordinary differential equations (ODEs) are the most widely used tool for modeling CAR T cell therapy, providing a framework to capture time-dependent dynamics of CAR T cells, tumor burden, cytokine release, and immune interactions.[11,11] For instance, Lotka-Volterra predator-prey dynamics have been adapted to describe interactions between CAR T cells and cancer cells (**Figure 1A**),[12] with extensions incorporating immunosuppression, CAR T cell differentiation, and bystander cell populations.[13–15]

Partial differential equations (PDEs) extend these models by adding spatial dependency, capturing how factors influencing CAR T performance diffuse through the tumor microenvironment and body.[16] Spatially dependent challenges with CAR T cells such as antigen heterogeneity, tumor infiltration, and immunosuppression can be modeled using PDEs.[17]

Mechanistic pharmacokinetic-pharmacodynamic (PK/PD) models, typically formulated with differential equations, capture CAR T cell kinetics—including rapid distribution, expansion, contraction, and persistence—and extrapolate dose-effect relationships for quantitative insights into therapeutic behavior.[18,19] These models differ from traditional PK/PD models because CAR T cells can dynamically change their population in response to tumor burden and other factors, unlike traditional drugs. These models must capture the growth rates and killing dynamics of CAR T cells and tumor cells, as well as CAR T cell differentiation into different cell states **(Figure 1B)**.[18]

***Stochastic Simulation Tools***

Stochastic modeling explicitly incorporates statistical noise into numerical simulations, allowing for each simulation to produce a different result based on random chance. Stochastic processes are used to model CAR T cell therapy to analyze and predict treatment outcomes, especially considering the inherent variability in cellular interactions and patient responses.[20]

There are several approaches to implement stochasticity into models of CAR T cells: it can be done using the Gillespie algorithm to model the chemical master equation (CME; a formalism to describe the time evolution of systems where the current random event only depends on the current state of the system), or stochastic differential equations (SDEs), where parameters of differential equations are randomly fluctuated at each time step. In the context of CAR T cell therapy, the CME can be applied to model the complex and often noisy interactions between CAR T cells and tumor cells, particularly when populations sizes are small and random fluctuations are significant. Incorporating stochasticity into CAR T dynamical models can help capture scenarios where deterministic approaches may fail to represent true biological variability, such as tumor elimination or relapse during CAR T therapy **(Figure 1C)**.[21]

***Rule-Based Tools***

Rule-based modeling simulates the cellular interactions or biochemical reaction networks using predefined rules derived from known biological mechanisms. This technique is especially effective for capturing the complexity of biochemical signaling systems in immunology, especially when the behaviors of single cells are somewhat predictable, but give rise to population-level emergent properties.[22,23]

Agent-based models (ABMs) define individual agents that follow a set of predefined rules based on their environmental conditions and interactions with other agents. For example, a CAR T cell "agent" can migrate, activate a killing response, or die depending on its context **(Figure 1D)**.[24] ABMs can also account for spatial dynamics (by allowing agents to transverse a two-dimensional or three dimensional domains) and stochasticity (by encoding decisions as probabilities).

In parallel, rule-based simulations of biochemical reaction networks have been applied to study the signaling dynamics of CARs.[25,26] Rather than enumerating every reaction, model developers specify domains and interaction rules, which can be translated into deterministic ODEs or stochastic simulations using platforms like BioNetGen.[27] These rule-based methods efficiently reduce the computational complexity of large signaling networks while preserving mechanistic fidelity.

### *Statistical and Machine Learning Tools*

Unlike hypothesis-driven tools for modeling CAR T cell dynamics, data-driven statistical learning tools seek to discover relationships between parameters and CAR T cell performance in a hypothesis-agnostic manner. These tools are well-suited to biological systems with complex, non-linear dynamics and numerous interacting and/or hidden variables, where underlying dynamics remain poorly understood.

Statistical learning techniques such as regression or decision trees are used to characterize the relationship between variables or make predictions about unseen data. For example, linear regression—a core statistical inference tool—can be used to quantify correlations between variables by fitting a line that minimizes the average error. Decision trees, on the other hand, can recursively partition data into increasingly homogeneous parts based on features of the data to minimize prediction error, resulting in an interpretable series of decision rules.

While statistical learning emphasizes interpretability, machine learning (ML) employs statistical learning to prioritize predictive accuracy. ML spans simple linear models tested on unseen data to deep neural networks capable of learning highly nonlinear relationships, such as links between CAR T costimulatory domains and phenotypes (**Figure 1E**).[28] While more complex architectures like deep learning can capture intricate patterns, they also require substantial training data, offer limited interpretability, and are prone to overfitting for small sample sizes.

Recently, large pre-trained deep learning models predicting protein structure, such as AlphaFold, ProteinMPNN, and RFdiffusion, have been applied to CAR T design, enabling binder prediction and protein engineering for next-generation therapies (**Figure 1F**).[29–31]

## REVIEW OF TRANSLATIONAL RESEARCH GOALS OF CAR T CELL MODELS

Next, we provide a strategic overview of modeling approaches for CAR T cell therapy, focusing on how recent advances in CAR T cell engineering open new avenues for model development. Rather than a traditional review, we map mathematical and computational models to key clinical and biological challenges, offering a framework for selecting appropriate tools (**Figure 2**). In this perspective, we also highlight opportunities to leverage clinical trial data for validation and integration, aiming to bridge computational and clinical communities and promote tailored modeling for improved CAR T design, evaluation, and personalization.

## Understanding and optimizing CAR T dosing regimens and treatment

The diversity of CAR designs, target patient populations, and preconditioning regimens across different cancer targets adds incredible complexity to optimizing CAR T cell treatment efficacy.[32,33] For example, an early trial targeting EGFRvIII in patients with recurrent glioblastoma multiforme (GBM)—an aggressive and treatment-resistant brain cancer—marked a milestone in extending CAR T therapy to solid tumors. However, its small cohort size, lack of a control group, and variability in patient characteristics such as tumor burden, immune status, and prior treatments limited the ability to rigorously evaluate safety and efficacy.[34] To address these limitations, mathematical and computational models can help predict effective dosing strategies and combinational treatments. In the following sections, we categorize these models by their increasing clinical realism and translational potential, showing how each one advances our understanding of CAR T optimization.

### *Characterizing CAR T–tumor population dynamics*

Efforts to model CAR T cells and cancer with Lotka-Volterra predator-prey dynamics effectively capture the behavior of *in vitro* CAR T killing assays, with CAR T cells—or "predators"—only able to grow in the presence of the tumor, and cancer cells—or "prey"—exhibiting logistic growth and CAR T-mediated death. The prototypic predator-prey model was developed by Sahoo et al., capturing glioma (prey)–CAR T (predator) cell dynamics *in vitro* and *in vivo*.[12]

Building on this framework, Li et al. demonstrated that explicitly modeling CAR T binding to glioma cells improves accuracy over Sahoo et al.'s quasi-steady-state binding approximation.[35] In an effort to discover the most accurate and simple models of CAR T and cancer cell population growth dynamics, others have applied machine learning powered algorithms to identify best fit governing equations (**Box 2**). These models establish a baseline approach for characterizing CAR T population dynamics and pave the way for more mechanistic models linking cellular behavior to molecular design.

### *Linking CAR T mechanisms to treatment outcomes*

Other mechanistic models capture antigen binding and CAR T cell activation. Singh et al. developed a physiologically based pharmacokinetic model (PBPK) incorporating CAR affinity, antigen density, and effector-to-target ratios to link molecular parameters with CAR T performance and biodistribution.[19] The results suggest that the peak concentration of CAR T cells in the blood depends more on the patient's tumor burden than on the administered dose, which is a fundamental difference between CAR T cells and traditional pharmaceuticals.

To capture the CAR T PK/PD profiles more accurately, Stein et al. developed a mixed-effects cellular kinetic model using phase II trial data of FDA approved CAR T therapy tisa-cel.[18] Their model incorporated effector and memory subsets of CAR T cells to reproduce expansion, rapid contraction and prolonged persistence. Their analysis showed that IL-6 blocking treatments for CRS did not affect CAR T expansion or decline, underscoring the value of modeling for clinical insights and trial design.[36]

Expanding clinical and phenotypic models, Hardiansyah et al. developed an early quantitate systems pharmacology model linking CAR T PK/PD to clinical data, cellular interactions, and cytokine signaling. The model reproduced patient cytokine dynamics (IL-6, IL-10, IFNγ), captured effector and memory differentiation, and quantified the interplay between CAR T dose, disease burden, and inflammatory cytokines, providing a validated platform to guide CAR T development and personalized dosing.[37]

Barros et al. extended these models to include immunosuppression and memory CAR T reactivation.[13] Similarly, Martínez-Rubio et al. incorporated healthy B cell populations and B cell development into their model to reproduce clinically observed phenomena such as B cell aplasia—a depletion of normal B cells that arises as an on-target side effect of anti-CD19 therapies.[38] These extended models offer *in silico* evaluation of dosing, infusion, and checkpoint inhibitor combinations, reflecting a shift toward more integrated representations of CAR T treatment response.

***Optimizing CAR T treatment strategies***

Beyond cellular kinetics, models also examine CAR T cell treatment alongside adjunct treatments.[39] CAR T cell therapy requires lymphodepletion (reduction in endogenous immune cells with chemotherapy), but regimens differ across treatments and may not be optimal. Owens et al. used a mechanistic population dynamics model, simulating tumor cells, endogenous immune cells, CAR T cells, and chemotherapy, and predicted that refining dosing schedules and timing between preconditioning and infusion can substantially improve treatment efficacy.[40]

Similarly, combinational therapeutic strategies involving CAR T cells with targeted radiation therapy (TRT) are being explored as potential strategies to mitigate adverse effects.[41,42] A recent study combined two previously published models: Sahoo et al.'s CARRGO model, and a model of TRT, to simulate combination treatment in multiple myeloma. The authors incorporated tumor radiosensitivity and CAR T dynamics, predicted superior survival when CAR T was administered before radiation, as TRT can impair CAR T cells.[43]

***Clinical Applications and Reflections***

In the context of optimizing CAR T dosing and treatment, future modeling efforts should prioritize integrating multi-modal clinical data and accounting for patient-specific covariates. Mixed-effects cellular kinetic models, such as those developed by Stein et al.,[18] could enable covariate-adjusted predictions of CAR T kinetics and outcomes, even in the absence of randomized control groups. Virtual control arms based on historical data could further improve trial power and interpretability. Importantly, approaches of this kind directly address the challenges highlighted in the EGFRvIII glioblastoma trial, pointing toward next-generation models that integrate mechanistic insight, clinical interpretability, and statistical inference to accelerate rational CAR T design across hematologic and solid tumors.

## Navigating Patient Variability

Despite its remarkable efficacy in some patients, a significant proportion of individuals receiving CAR T treatment experience suboptimal responses, relapses, or treatment failure due to biological and clinical variability. One representative early clinical trial that illustrates this variability is the ELIANA trial of CD19-directed tisa-cel in B-ALL, which showed dramatic long-lasting remissions in 81%, patients within 3 months, but nonresponse in others.[44] Such divergent outcomes highlight the critical need to understand and anticipate variability in patient responses. Mathematical and computational models are increasingly being employed to deconstruct and simulate CAR T–tumor–immune interactions, advancing from mechanistic to data-driven frameworks to better capture heterogeneity and improve translational predictive power.

### *Capturing patient variability with stochasticity*

Stochastic models help investigate the variability of patient responses to CAR T treatment. Kimmel et al. used a hybrid deterministic and stochastic model to explore patient variability in CAR T cell by treating tumor cure as a stochastic event, where tumor cells enter a stochastic regime (where the Gillespie algorithm is used to model the chemical master equation) when depleted below a count of 100.[21] This study indicated that stochastic cure alone cannot explain patient heterogeneity.

Hoang et al. extended this model and adapted it to a stochastic differential equation framework, introducing different intensities of white noise into each parameter values to capture patient variability. Their model of transient dynamics could be used in clinical decision making to time interventions and predict outcomes.[45]

### *Uncovering drivers of variability*

A clinical data-driven ODE model using data from 209 B-cell acute lymphoblastic leukemia (B-ALL) patients modeled heterogenous patient responses including complete responses, disease progression, and both CD19+ and CD19– relapse dynamics. Their model, which included mutation rate of the cancer and bystander killing effects of CD19– cells, was able to forecast patient outcomes based on early CAR T cell expansion with over 74% accuracy. This *in silico* approach was validated using virtual patient cohorts and highlights the promise of clinical response prediction models using simulated patient data.[46]

Paixão et al. developed a deterministic ODE-based model to capture the inter-patient variability in CAR T kinetics using a customizable function governing CAR T cell expansion that can be adjusted to each patient in trial data across three cancer types (DLBCL, ALL, and CLL). They found that variations in CAR T cell phenotypes (such as memory, exhausted, and effector subsets) were insufficient to explain the observed differences in PK/PD profiles.[47]

Building on concepts from previous studies,[21,38,47] Kirouac et al. developed a PK-PD model which describes transitions of CAR T cells from effector to exhausted and memory-like states through an antigen-driven toggle switch.[48] Similar to the approach taken by Paixão et al.,[47] the authors showed that differences in patient response to CAR T therapy could not be fully explained by standard immunophenotyping (e.g., measuring the proportion of memory and exhausted CAR T cells at the time of infusion). To address this gap, they extended their framework by applying a machine learning classifier to bulk RNA sequencing data to stratify response groups (complete, partial, or none) based on transcriptomic signatures. This integrative approach combining PK-PD modelling with transcriptomic data could be used to guide future CAR T design, highlighting the value of coupling mechanistic models with data-driven tools.

***Forecasting patient outcomes***

Other modeling approaches seek to draw correlations between measurements and treatment outcomes without regard to mechanism. Finney et al. used the machine learning technique called classification and regression tree (CART) analysis to examine how features of the engineered CAR T cells prepared for infusion relate to treatment outcomes.[49] They found that measuring specific biomarkers in patients' CD8+ T cells before genetic modification could help predict individual responses to therapy, achieving a coefficient of determination ($r^2$) of 0.636.

Machine learning models trained on cytometry and medical imaging (such as PET and CT scans) have been used to predict how patients will respond to CAR T therapy. Notably, one model achieved a high prediction accuracy, with an area under the curve (AUC) of 0.91 out of 1.[50] Similarly, Jak et al. trained a machine learning model using pre-treatment PET and CT scan images from patients with DLBCL who were scheduled to receive the CAR T cell therapy. The model aimed to predict patient outcomes based on these imaging features.[51] These models may become critical for patient selection.

***Clinical Applications and Reflections***

Together, these studies converge on the insight that heterogeneity in treatment outcomes arises from a complex interplay of tumor burden, immune environment, CAR T cell product characteristics, and patient-specific physiology—factors that cannot be fully explained by any single framework. Integrative approaches that combine mechanistic understanding with data-driven inference are therefore increasingly necessary to inform clinical translation and personalized treatment planning. Importantly, stochastic and stochastic differential equation models provide a means to simulate uncertainty in tumor clearance and explore variability in treatment durability. Embedding such models into trial design—particularly in small, heterogeneous cohorts like those of the ELIANA study—could enable predictive, adaptive strategies that reduce relapse and improve the consistency of CAR T outcomes.

## Predicting and Managing Adverse Effects

The regulatory evaluation of CAR T cell therapies is challenging given their variability and risk of severe toxicities such as cytokine release syndrome (CRS) and Immune Effector Cell-Associated Neurotoxicity Syndrome (ICANS).[52,53] One prevalent foundational clinical trial where many patients experienced adverse effects were the JULIET trial for relapsed/refractory DLBCL.[54,55] In JULIET, tisa-cel induced CRS in over half of patients, with about one-fifth experiencing severe CRS, while neurotoxicity occurred in a smaller subset. This trial illustrates the need to improve our understanding of the mechanisms behind these severe side effects. Mathematical models can predict variability in adverse event onset, aiding patient selection, therapeutic dosing strategies, and the design of safer therapies.[40]

### *Predicting CRS onset*

Building on Paixão et al.'s model of multiphasic CAR T cell kinetics, Santurio et al. extended the framework to incorporate interactions with macrophages, an innate immune cell type thought to play a key role in the onset of CRS.[10] They modeled three pathways of macrophage activation: (1) cytokine signaling triggered by activated CAR T cells, (2) the CD40–CD40L costimulatory axis, and (3) molecular signals known as damage-associated molecular patterns (DAMPs) released from dying tumor cells. By removing each activation pathway in simulations, they evaluated how different therapeutic interventions could reduce peak cytokine levels linked to CRS and found that the CD40–CD40L axis had the greatest impact.

### *Predicting clinical toxicities*

Many models use clinical measurements to predict the onset of adverse events, parallel to the previously discussed models that use similar methods to predict the outcome of CAR T treatment. Teachey et al. measured cytokines and clinical biomarkers CAR T-treated patients and used forward-selected logistic regression to identify a three-cytokine signature (IFNγ, sgp130, sIL1RA) that predicted severe CRS, enabling potential early intervention.[56]

In a transcriptomic-centered approach, Good et al. used cytometry time-of-flight (CyTOF) and single-cell RNA sequencing data from B-ALL patients to find CAR T phenotypic subsets predictive of response and ICANS severity. Their models achieved ~75% accuracy for treatment response and ~77% for severe ICANS, underscoring the value of integrating clinical, cytokine, and transcriptomic biomarkers with modeling to anticipate patient outcomes.

Alongside outcome prediction, clinical and imaging data from PET/CT scans are being used to forecast CAR T toxicities. Ferrer Lores et al. developed a logistic regression model to predict the risk of CAR T-related neurotoxicity using clinical data and pre-treatment PET or CT scan features, including tumor size and metabolic activity.[57] In 29 patients with DLBCL and PMBCL, the model accurately predicted ICANS incidence (AUC

= 0.971), highlighting the potential of imaging-based predictors to guide treatment strategies and prophylactic interventions.

### *Clinical Applications and Reflections*

Together, these modeling frameworks highlight the complexity of treatment-related toxicities and demonstrate the potential of quantitative models to guide clinical risk stratification, inform dosing strategies, and support early intervention. These approaches represent critical steps toward personalized CAR T administration that balances efficacy with safety.

Integrating recent modeling frameworks into clinical decision making could help to improve patient safety, such as including immune–CAR T interactions to simulate patient-specific cytokine surges, which could allow proactive testing of interventions *in silico* before treatment.[10] In addition to imaging features currently used to identify CRS risk,[54,55] pre-treatment blood-based biomarkers, such as those by Teachey et al. and Good et al., could serve as non-invasive and preemptive methods to inform patient selection for CAR T therapy based on their risk of severe adverse effects. Looking ahead, integrated models that combine mechanistic biology, biomarkers, and imaging data are well positioned to guide patient management and accelerate the design of safer, more effective CAR T therapies.

## Designing and Controlling CAR T cell phenotype

As “living therapeutics,” CAR T cells sometimes exhibit unexpected or unexplained behavior in *in vitro*, preclinical, and clinical settings. CAR T cells can fail to proliferate or proliferate too much, become exhausted and lose their ability to kill cancer cells over time, adopt immunosuppressive phenotypes such as Tregs, or even develop into T-cell based secondary malignancies. One trial that encountered this roadblock was the ZUMA-1 trial of axi-cel for refractory large B-cell lymphoma, where more than half of the patients eventually relapsed post-CAR T infusion.[54] As subsequent ML-based analyses of these and other patients treated with axi-cel revealed,[58] a major predictor of relapse was the development of CAR Treg cells, where the infusion product developed immunosuppressive characteristics. This study underscores that designing safer and more effective CAR T cell therapies will involve improving our control over their phenotypic differentiation. In this section, we highlight modeling efforts that start by exploring fundamental cellular mechanisms, then advance to systems-level simulations of CAR T behavior and ultimately integrate data-driven methods for design and clinical application.

### *Regulating CAR T activation and exhaustion*

Many studies have modeled the intracellular signals triggered by the chimeric antigen receptor (CAR) that influence CAR T cell behavior. Harris et al. used a kinetic proofreading model to compare CARs with natural T cell receptors (TCRs), which recognize antigens during immune responses. They found that CARs signal less efficiently than TCRs, likely due to slower or less effective phosphorylation of key signaling proteins.[59]

Others have employed mechanistic models to understand how costimulatory domains can improve T cell activation. Rohrs et al.[25] tracked Lck (a key kinase that initiates T cell signaling) -mediated phosphorylation of CAR proteins with phospho-proteomic mass spectrometry. They fit the data to ODE-based models, identifying their competitive inhibition model as the best match. They later developed a comprehensive, experimentally validated model of CAR signaling that incorporated antigen binding and MAPK/ERK pathway activation, offering one of the first detailed intracellular frameworks of CAR T cell activation.[60]

Tserunyan and Finley also studied how the 4-1BB costimulatory domain (which promotes T cell survival) affects CAR T cell signaling.[61] Using a mechanistic model of 4-1BB–induced NFκB activation, they applied information theory to test antigen detection under noisy conditions,[62] and predicted that NEMO overexpression could enhance antigen sensitivity. Cess and Finley used partial least squares (PLS) analysis on Monte Carlo simulations of their earlier model, identifying that only proteins closely linked to the CAR receptor and the MAPK pathway strongly influenced T cell behavior.[26]

Others have used Boolean models to decode the regulatory logic of T cell exhaustion.[63] Bolouri et al. mapped a CD8+ T cell signaling network with two opposing programs: pro-memory/proliferative (PP) and effector/exhausted (EE), driven by mutual gene inhibition. Their model showed exhaustion arises from network feedback loops, not single genes, and predicted that blocking EZH2 prolongs the PP state without raising exhaustion markers, which has since been validated experimentally.[64] More broadly, such models reveal regulatory circuits controlling activation and exhaustion.

### *Optimizing CAR T function*

Machine learning (ML) is increasingly used to optimize different genetic and signaling designs of CAR T cells. Daniels et al. built a CAR library with diverse signaling motifs and tested their cytotoxicity and stemness by surface marker staining and flow cytometry. They subsequently trained an ML model to predict untested designs that maximize cytotoxicity and stemness. yielding a new CAR costimulatory domain design that outperformed the standard anti-CD19 4-1BBζ construct *in vitro* and *in vivo*.[28]

Qiu et al. developed *CAR Toner*, an ML model predicting tonic signaling from positively charged patches on CAR surfaces. It identified point mutations that reduce tonic signaling, experimentally validated to lower exhaustion markers and enhance killing, demonstrating ML's potential for CAR functional optimization.[65,66]

### *Clinical Applications and Reflections*

A central insight is that CAR T cell phenotype (such as cytotoxicity, persistence, memory formation, and exhaustion) arises from both engineered features (such as costimulatory domains or scFv architecture) and emergent host immune dynamics. Mechanistic models identify signaling-level design knobs, while data-driven and machine learning approaches explore large design spaces and uncover hidden functional patterns. Motivated by clinical outcomes such as those in the ZUMA-1 trial as shown by Good et al., future work should focus on integrate mathematical and computational models into the design pipeline of CAR T cells. For example, expanding on existing PK/PD models to include differentiation into CAR Treg cells could enable simulation of Treg-depleted or memory-enriched products and optimized dosing. Integrating signaling models, regulatory network simulations, and ML-guided engineering with clinical phenotype data enables rational tuning of therapies, maximizing durable responses while minimizing adverse effects, shifting development from empirical observation to predictive, mechanism-based design.

## Addressing heterogeneous and evolving antigen expression

A major challenge in CAR T cell engineering is discovering rare antigens present exclusively on all cancer cells but absent from healthy cells to prevent what's termed as on-target, off-tumor toxicity. Tumors also heterogeneously express antigens, making finding a single target that is expressed on every cancer cell difficult or impossible. Furthermore, many tumors have high mutation rates, resulting in dynamic and unstable antigen expression levels that allow them to evade targeting. One illustrative example that impacted the field was a case study where a patient with metastatic colon cancer was treated with HER2-directed CAR T cells.[67] Following treatment, the patient experienced fatal lung damage, likely from low-level HER2 expression on normal lung tissue, underscoring risks of non-tumor exclusive targets. In this section we present mathematical and computational models of antigen expression and escape, with each step deepening biological insight and clinical relevance.

### *Uncovering mechanisms of antigen escape*

Santurio et al. developed a mechanistic model to explore resistance in CD19-targeted CAR T therapy, focusing on how transient and antigen loss contribute to relapse. Tracking antigen-positive and -negative tumor subpopulations plus effector and memory CAR T cells, the model fit data from 18 patients and reproduced complete responses, antigen-positive relapse, and antigen-negative relapse, with mutation-driven antigen loss best explaining resistance.[16] Others have applied ABMs to study the spatiotemporal dynamics of heterogenous antigen expression, hypothesizing that antigen-negative cells can shield antigen positive cells from recognition, thereby reducing the effectiveness of CAR T cell therapy.[68]

### *Designing antigen-targeting strategies*

ABMs have also been applied to the issue of on-target off-tumor toxicity. One notable example was the CARCADE model published by Prybutok et al.[24] In this model, CAR T cell agents were represented as one of ten states, including desirable states where CAR T cells remain active, mobile, and capable of expanding or killing tumors, as well as undesirable states where CAR T cells lose function, die, or become unresponsive. Their study developed the hypothesis high dosages of low-affinity CAR T cells could provide a potential treatment window to maximize CAR T cell killing while limiting toxicity, which warrants experimental validation.

Some studies have applied mechanistic ODE models to measure other effects, such as Santurio et al.'s study exploring off-target toxicity, where CAR T cells target antigen-low cells such as healthy glial cells in glioblastoma.[69] Kara et al. also used a mechanistic ODE model to explore bystander effects, where activated immune cells near the tumor go on to kill healthy cells that don't express antigen.[70] León-Triana et al. created a model of dual-targeting CAR T cells that target both a tumor-associated antigen (IL13Rα2 on glioblastoma) and an abundant off-tumor antigen (CD19), showing that the dual-targeted CAR could control the tumor over a longer period of time.[14]

Finally, A series of recent studies[29–31] applied established ML models, including AlphaFold2/3,[71,72] ProteinMPNN,[73] and RFDiffusion[74] to design *de novo* binders for peptide-MHC complexes, that can be used to replace the extracellular single-chain variable fragment (scFv) portion of the CAR. These models generated binder designs which were then experimentally screened by binding assays, which measured specificity and affinity. All three studies produced functional CAR T cells validated *in vitro*, demonstrating ML pipelines can both discover novel antigen binders and improve specificity and affinity of existing CAR designs.

***Optimizing CAR T specificity***

Complementary approaches have integrated intracellular signaling with population-level dynamics to evaluate how CAR design features affect targeting specificity. One such effort is the PASCAR model developed by Rajakaruna et al.,[75] which uses a multi-scale approach to evaluate how antigen heterogeneity affects the ability of CAR T cells to selectively target tumor cells while sparing healthy tissue. At the molecular level, the model incorporates kinetic proofreading, a stepwise signaling process in which only strong, sustained antigen binding triggers full activation,[76] while weaker interactions fail to proceed.[77] PASCAR couples this ODE-based molecular-scale module to a population-scale model of the interactions between CAR T cells and diverse target cell populations, accounting for variation in both CAR expression and antigen abundance. This multi-scale approach enables quantitative prediction of *in vitro* outcomes, including tumor cell killing rates and off-target toxicity. Notably, the model also applies a multi-objective optimization approach, known as Pareto optimization, to evaluate trade-offs between competing goals of maximizing tumor clearance while minimizing off-target effects. Through this analysis, PASCAR identifies design parameters, such as specific antigen binding affinities, that achieve an optimal balance between these goals. By combining mechanistic signaling insights with clinically relevant performance criteria, this work exemplifies how modeling can guide rational CAR design in the face of phenotypic and antigenic heterogeneity.

***Clinical Applications and Reflections***

Collectively, the modeling efforts underscore the essential role of mathematical and computational approaches in addressing antigen heterogeneity, off-tumor toxicity, and antigen-negative relapse in CAR T therapy. A key insight is that models incorporating continuous variation in antigen expression or CAR expression reveal how subtle differences in antigen density can flip outcomes between effective tumor clearance and unintended damage to healthy tissue. Modeling frameworks such as PASCAR, which incorporates kinetic proofreading and simulates CAR–ligand interactions across heterogeneous populations, or agent-based models like CARCADE, could help anticipate adverse outcomes, such as the case study of HER2 CAR T cell-induced fatality, by identifying affinity ranges or combinatorial strategies that minimize off-tumor reactivity.

## Infiltrating and surviving within the tumor microenvironment

One of the major challenges in adapting CAR T cell therapy to solid tumors is enabling the engineered cells to effectively infiltrate and survive within the immunosuppressive tumor microenvironment (TME). Solid tumors often consist of dense, poorly vascularized tissue that contain suppressive immune cell populations—including myeloid-derived suppressor cells (MDSCs), tumor-associated macrophages (TAMs), and regulatory T cells (Tregs) that both limit T cell trafficking into the tumor and impair the function of any CAR T cells that manage to infiltrate the tumor.

One such clinical trial example that faced this roadblock was a phase I trial of prostrate-specific membrane antigen (PSMA)-directed CAR T cells that were modified to be insensitive to TGF-β, (a major immunosuppressive cytokine) in metastatic castration-resistant prostate cancer (mCRPC).[78] These "armored" CAR T cells were able to successfully infiltrate the tumor in many patients, however digital spatial profiling revealed that there was upregulated immunosuppressive molecules in the T cell-rich tumor areas that may have comprised therapeutic function, leading to overall poor efficacy in the trial. In the following paragraphs, we highlight modeling efforts that begin by characterizing physical barriers to CAR T cell trafficking, then incorporate immunosuppressive interactions within the tumor microenvironment, and ultimately simulate complex checkpoint signaling dynamics, with each step deepening biological insight and informing clinical strategies.

### *Characterizing CAR T infiltration and delivery*

Partial differential equation (PDE)-based reaction-diffusion models have been used to capture the spatial aspects of CAR T cell delivery and tumor penetration. Owens et al. modeled intratumoral (injection into the tumor mass) versus intracavitary (injection into surrounding fluid-filled spaces) *in vivo* delivery of CAR T cells towards tumors with varying density and growth rates, incorporating effector and exhausted CAR T states. The model predicted intracavitary delivery alone could not induce complete response, though further refinement is needed to reflect preclinical and clinical successes with systemic and intracavitary CAR T administration.[79] Bordel-Vozmediano et al. similarly used an ABM to explore the spatial dynamics of CAR T killing of dense and sparse tumors *in vitro*.[80] Predictably, they found that CAR T cells are more effective at killing and proliferating within sparse tumors.

### *Characterizing immunosuppression*

Rodrigues et al. reported an ODE-based model that captured the interactions between CAR T cells, melanoma tumor cells, and immunosuppressive tumor-associated macrophages (TAMs).[81] Their equations incorporated terms where TAMs inhibit CAR T cell expansion and promote tumor growth. Although not yet parameterized with experimental data, the model offers a conceptual foundation for future extensions that could incorporate the effect of immunosuppressive macrophages on CAR T cells.

Just as spatial heterogeneity in antigen expression shapes CAR T activity, the spatial distribution of immunosuppressive factors can affect CAR T performance in the tumor microenvironment. ABMs and PDEs could be a promising method to explore these dynamics through investigating how different properties of tumors, including mutational burden and antigen strength, affect the spatial distribution of immunosuppressive properties, such as PDL1 expression.[82] Others have created PDE models of TAM subtypes to model the effects of different immunotherapeutic strategies (e.g. TAM ablation vs. TAM re-education) on tumor growth.[83] These models could be expanded to include the effect of CAR T cells on PDL1 expression patterns and vice versa, which could help to explain the dynamics of immunosuppression of CAR T cells and their potential synergy with immune checkpoint inhibitor therapy.

### *Clinical Applications and Reflections*

To better characterize and mitigate the effects of TME immunosuppression, Rodrigues et al.'s model of the effects of immunosuppressive TAM populations could be parametrized with patient data from trials like the PSMA-directed CAR T trial, and modified to include modeling of specific immunosuppressive factors such as TGF-β. Then these models could be applied to test out different armored CAR T cell designs, such as the TGF-β insensitivity from this trail, to observe if removing this axis of immunosuppression could significantly improve the therapy. Overall, there is an opportunity to further develop more models of tumor infiltration and immunosuppression of CAR T cells to better understand and overcome these major barriers to clinical success.

# DISCUSSION

Modeling CAR T cells has provided valuable insights in understanding CAR T behavior, revealing key dynamics that inform improved CAR-designs, superior treatment regimens, and tools for predicting patient response and adverse effects. Despite their considerable potential in improving CAR T cell therapy, mathematical models remain underutilized by the field at large. We believe that new opportunities in CAR T cell modeling are emerging, and below are perspectives on some of the most promising novel directions and applications of modeling in advancing the translational potential CAR T cell therapy.

## The paradigm shift in data quality and quantity

One of the central challenges to many of the studies covered in this perspective was the inability to fully constrain model parameters due to a lack of supporting data. For these models to serve the role of advancing our understanding of CAR T cell therapy and its clinical translation, it is essential to generate diverse, high-resolution datasets that can reduce parameter uncertainty and sharpen predictions.

Recent advances are addressing this need, with clinical trial data, high-throughput *in vitro* and *in vivo* screens, and genomic profiling technologies such as single-cell and spatial transcriptomics increasing in quantity and quality. Further, these datasets are increasingly being made public through open science efforts,[84–87] thereby increasing their use in model development.[88] Below, we outline the advances in *in vitro*, *in vivo,* and clinical data quantity and resolution that we believe will greatly accelerate CAR T modeling efforts if incorporated properly.

Larger scale 96-well arrayed *in vitro* screens, powered by automated live cell imaging platforms and flow cytometry, enable screening hundreds of different CAR T cell designs and yield diverse phenotypic readouts (proliferation, killing, markers of differentiation/exhaustion) across varied conditions.[51] These time-resolved datasets under different effector-to-target ratios, immunosuppressive cues, or presence of other relevant cell populations help constrain dynamic models. This level of parallelization in the cell engineering process has become the new standard, similar to how high throughput screening (HTS) of small molecules has become the industry standard and a highly productive way to generate new candidates for pharmaceuticals.[93]

Another datatype increasing in resolution and scale that provides new opportunities for CAR T cell modeling is transcriptomics, which measures genome-wide gene expression levels using RNA sequencing. Gene expression in populations cells can be measured either in pooled through bulk RNA sequencing, at the single-cell level through single-cell RNA sequencing, or with single-cell or spot-level spatial resolution through spatial transcriptomics. Single-cell transcriptomics of CAR T infusion products or blood and tumor samples can initialize ABMs with more detailed states than current models based on binary parameters such as cytotoxicity, T cell subtype, or proliferation. It also provides rich training datasets for machine learning models, for example to predict phenotypic effects of genetic perturbations when conducted on gene knockout populations of CAR T cells,[89] or to identify transcriptomic differences in CAR T cells with different behaviors, such as tumor infiltrators or exhausted, immunosuppressed, or cytotoxic populations.

Another application of RNA sequencing is to enable pooled screening, where hundreds to thousands of distinct CAR T cell variants are cultured together under functional challenges like tumor cells or immunosuppressive factors.[90] The relative fitness of each variant is tracked via its unique genetic barcode using sequencing, or measurements of surface markers or proliferation dyes through flow cytometry. This generates population enrichment datasets useful for training ML models to predict desirable CAR T phenotypes.

Furthermore, spatial transcriptomics adds neighborhood-level context to CAR T cell dynamics by mapping local antigen density, stromal barriers, and immunosuppressive cues. For example, *in vivo* knockout perturbations of select genes by Dhainaut et al. and Binan et al. in tumor cells was conducted and their effect on immune cell infiltration[91] or gene expression of neighboring T cells[92] was measured by spatial transcriptomics. Extending this to CAR T cells screening gene knockouts could reveal therapy-amplifying effects on nearby tumor cells or immune cells, creating a roadmap for genetically improved CAR T cell therapy. This could generate high resolution training data mapping genes to CAR T performance, enabling calibration of PDE or hybrid PDE–ABM terms for chemotaxis, diffusion, extravasation, infiltration, and contact-dependent engagement. Moreover, ML algorithms could be trained to predict effects of combinatorial gene edits across otherwise experimentally infeasible design spaces.

Another promising technology that could provide new opportunities for CAR T modeling is intravital imaging, a technique where fluorescent cells can be imaged via multi-photon microscopy within a live animal by implanting a small imaging window. Intravital imaging provides direct, time-resolved measurements of CAR T cell behavior inside the tumor microenvironment.[93] Unlike spatial transcriptomics, these technologies allow measurements of temporal dynamics in addition to the spatial dimensions of CAR T therapy—capturing CAR T infiltration, migration velocity, search patterns, and dwell times at the tumor-immune interface.[94–96] These data are essential for ABMs or PDE frameworks, where cell motility coefficients and chemotaxis parameters must be specified. Additionally, the observed durations of contact between CAR T cells and tumor cells can provide crucial data for determining per-contact killing probabilities within ABM frameworks. Recent advances in parameterizing ABMs with tumor images can help to empower these types of studies.[97]

Large machine learning models have already been demonstrated to provide immense utility when enough high-quality data has been generated to train them. The abundance of solved protein structures—primarily measured through X-ray crystallography and cryo-electron microscopy—have enabled large-scale machine learning models such as AlphaFold, ESMFold,[98] and ProteinMPNN to accurately map sequence to structure, while models like RFdiffusion can generate entirely *de novo* structures. In parallel, large language models (LLMs) trained on extensive sequencing data have produced DNA language models capable of designing new proteins, such as novel Cas endonucleases and transposases.[99] These modern LLM-based approaches could be applied to dissect signaling cascades and generate *de novo* CAR binders, signaling domains, and synthetic cytokines or cytokine receptors to fine-tune CAR T cell phenotype. Together with advances in single-cell multi-omics and large-scale clinical datasets, these tools are driving a paradigm shift in how CAR T therapies are modeled and optimized.

## The potential of hybrid models for multi-scale phenomena

Understanding and controlling CAR T cell behavior requires integrating knowledge of biomolecular phenomena across vast length and time scales. The chimeric antigen receptor itself works at the molecular level, binding to its antigen target and transmitting signals to the cell through phosphorylation cascades, that can occur in seconds to minutes. These signals then cause transcriptional changes in the CAR T cells in a matter of hours, that led to changes at the phenotypic level within days. Many of these proteins include communication factors that change the cell's interactions with cancer cells—such as inducing a killing response or exhaustion—or its interaction with other immune cells and bystander cells. These changes lead to highly nonlinear population shifts on the scale of days, weeks and months. Ultimately, cure and progression are defined by cancer growth at the tissue scale, including relapse dynamics that can take several years. Many studies still rely on one modeling framework (e.g., ODEs or agent-based models) to approximate dynamics across all levels. However, a single modeling approach rarely captures the complexity of CAR T cell dynamics in its entirety, limiting mechanistic insight and translational predictive power.

Some models attempt to bridge these spatiotemporal scales, such as the PASCAR model which unifies cell signaling scale dynamics with population dynamics by incorporating details of CAR T signaling implicitly within population-scale parameters.[75] However, some modeling formalisms may be particularly useful for one of these scales and not others. Since a single formalism may not suit all scales, hybrid models offer a powerful strategy, combining complementary frameworks to leverage strengths while mitigating their limitations. In other immunological models, mechanistic ODE-based or Boolean-based models of intracellular signaling have been combined with population-level ABMs to link cell-level events to patient outcomes.[101,102] This hybrid and multi-scale approach can link cell-level signaling events to patient-level outcomes, but their use in CAR T remains limited, offering opportunities for multidisciplinary work.

Another advantage of hybrid approaches is to separate population-level characterization from cell-signaling level implementation, a key step in linking CAR T dynamics across scales to guide therapy design. Some seek to engineer CAR T cells to exhibit desired properties, while others focus on identifying which features are most desirable or predictive of complete responses. These fundamentally distinct questions are also separated by spatiotemporal scale, with CAR T cell design depending on mechanistic understanding of signaling and transcription driven by synthetic genetic components, while population-level efficacy relies on their ability to grow, migrate through the body, and persist long-term. Hybrid multi-scale models can constrain tumor–CAR T interactions with experimental population data (e.g., tumor burden kinetics, cytokine curves), while independently designing or testing intracellular signaling via mechanistic models. Such separation may reduce overfitting, improve interpretability, and enable systematic testing of genetic circuit designs at the signaling level before integrating them into population-scale tumor–CAR T models.

Finally, hybrid models also unify diverse data, with transcriptomics refining CAR T state transitions (effector to memory to exhausted) and tumor regression data anchoring system dynamics. Kirouac et al. demonstrated this by combining a PK/PD tumor model with a transcriptomic ML classifier, capturing both mechanistic detail and patient heterogeneity.

## Expanding immune cell therapy models to include emerging cell types

In addition to CAR T cell therapy, CAR Macrophage (CAR MΦ) and CAR Natural Killer (CAR NK) therapy are emerging cell therapies that exploit the unique strengths of macrophages and NK cells as a chassis for the chimeric antigen receptor **(Table 3)**, offering alternatives for tumors less responsive to CAR T.[103,104] The integration of CAR MΦ and CAR NK therapies with existing immunotherapies provides new avenues for treating a broader spectrum of cancers, including those resistant to conventional treatments.[105] The first generations of these technologies have already made it to clinical trials and are expected to reach broader clinical application in near future.[106–108] Despite their promises, mathematical modeling has seldom been applied to these newer cell therapy platforms, representing an important opportunity for future research.

Modeling CAR MΦs requires major revisions to CAR T frameworks, as their PK/PD dynamics and biodistribution remain incompletely understood. Notably lacking proliferative capacity, they exhibit distinct engraftment dynamics, making their PK modeling fundamentally different from the expansion–contraction–persistence phases of CAR T cells.[109] Additionally, natural macrophages have strong tumor-homing capabilities, but it remains unclear whether CAR MΦs have a superior ability to enter the tumor microenvironment when compared to CAR T cells.[110] PK/PD models could be adapted to guide CAR MΦ dosing and delivery, potentially at lower doses due to strong tumor-homing. Unlike the rigid phenotypes of CAR T cells, macrophages exhibit plastic polarization between pro- and anti-inflammatory states. Models should incorporate this polarization to capture state emergence in CAR MΦs.[111–113] This could inform if certain genetic modifications could polarize or stabilize CAR MΦs to desirable phenotypes, and also capture how these states reshape the tumor microenvironment in either beneficial or detrimental ways. Finally, unlike CAR Ts, CAR MΦs kill via phagocytosis or cytokine-mediated mechanisms.[114,115] Capturing these different killing mechanisms and dynamics within models will be essential to inform future genetic design decisions.

CAR NK cells present different modeling opportunities. Beyond CAR recognition, they detect diverse tumor ligands, making population models useful to assess relapse risk from antigen loss compared to CAR T cells. Additionally, given their less varied phenotypic states compared to CAR T or CAR MΦ cells, they are well-suited for agent-based and ODE models, with fewer parameters and more interpretable dynamics. Their PK/PD dynamics also differ significantly from CAR T cells, notably exhibiting significantly reduced persistence. To address this, modeling can explore the effects of multiple infusions—a strategy used in CAR T cell therapies and increasingly relevant to CAR NK clinical trials.[116] Another advantage of NK cells is their ability to be derived from diverse sources, since allogenic NK cells are not rejected by the host and do not cause graft-versus-host-disease enabling them to be sourced from cell lines, primary cells, cord blood, or stem cells. Parameterizing these sources could reveal distinct phenotypic traits of CAR NK products.

Early efforts to model CAR NK cells and CAR MΦs have already emerged. Arabameri and Arab explored ODEs and ABMs to study CAR NK cells and triple-negative breast cancer (TNBC) interactions, examining persistence, therapeutic efficacy, Treg suppression and patient outcomes.[117] PDE- models have also been applied to CAR MΦs for treating SARS-CoV-2 infection, capturing polarization by virions and T cells to identify signaling domains that reduce inflammation while boosting phagocytosis.[118] These models serve as foundations for future models to further develop accurate predictions and mechanistic understanding of these emerging therapies.

As these therapies become more prominent in the clinic, modeling will be critical to dissect complex behaviors and guide optimization. A promising direction is designing synthetic immune systems combining engineered cell types, leveraging synergy between effector and modulatory functions to boost efficacy and reduce toxicity. Although early attempts show mixed outcomes,[110,119] modeling could help explore and optimize these multi-cellular strategies.

## Framing CAR T cell engineering with concepts from control theory

A fundamental challenge of CAR T cell therapy is achieving therapeutic robustness and control, i.e., the ability to sustain effective function despite biological variability and clinical uncertainties. As “living therapeutics,” CAR T cells must be engineered to self-regulate through molecular feedback mechanisms to adapt to varying conditions or to include externally triggered shutdown mechanisms that guard against malfunction. Further, CAR T cell treatments strategies must be designed through dosing regimens and strategies that balance and optimize multiple goals including tumor clearance, patient safety, and prevention of relapse, which can be explicitly framed as an optimal control problem. Thus, there are multiple scales ranging from the intracellular- to the patient-level where control principles can be leveraged to improve the safety, efficacy, and robustness of CAR T cell therapy.

As mathematical models of CAR T cells advance, control-theoretic analysis helps to design therapies robust to variability across scales, from molecular-level stresses such as physical barriers, T cell exhaustion, and immunosuppressive signaling to patient-level factors such as tumor heterogeneity, differences in tumor burden, and comorbidities. Just as control engineering has been essential for ensuring the safety and reliability of engineered systems such as aircraft flight control and automotive braking, it may be equally critical to integrate control-theoretic principles into the design of living therapeutics to achieve safe and reliable therapeutic behaviors.

In CAR T cell engineering, the first synthetic feedback mechanisms have already been designed. For example, Fedorov et al. engineered CAR T cells with inhibitory chimeric antigen receptors (iCARs) that deliver suppressive signals upon engagement with healthy tissue antigens.[120] This design functions as a synthetic negative feedback loop, dynamically self-regulating T cell activity to prevent off-target toxicity and preserve therapeutic function. The approach parallels feedback controllers in engineering, where system outputs are continuously monitored and adjusted through self-regulation to maintain stability.

Some strategies emphasize external rather than internal feedback control. Guercio et al. developed a "safety switch" system in CAR T cells using inducible caspase-9 (iCasp9), which allows for the controlled elimination of CAR T cells in the event of severe toxicity.[121] Additionally, Weber et al. demonstrated that externally induced “rest” periods using drug-regulatable domains can reset exhausted CAR T cells and restore their function.[122] These approaches parallel override controllers in engineering, where safety mechanisms are built-in to shut a system down in the event of failure or to pause and restart its activity to maintain stability. Other designs emphasize conditional control. For example, Hernandez-Lopez et al. engineered a two-step synNotch-to-CAR system that filters for high antigen levels before activating a strong CAR response, creating a switch-like mechanism that targets tumor cells while sparing normal tissue.[123] The approach parallels feedforward controllers in engineering, where input conditions are used to preemptively adjust downstream responses without requiring continuous feedback monitoring.

Even though these studies conceptually reflect ideas from control theory, their outcomes could be strengthened by incorporating formal control-theoretic modeling. Control theory is a useful way to frame CAR T design challenges, and through rigorous mathematics, can guide design choices, generate testable hypotheses, and improve phenotype reliability. One notable study that formalizes the control-theoretic metaphor in design of engineered cell systems was conducted by Ma et al., who developed a dual positive–negative (paradoxical) feedback design to regulate mammalian cell populations as a prototype for therapeutic applications. Inspired by natural T cell regulation,[124] this design produced a biphasic response that stabilized population size and improved resistance to mutational escape beyond simple negative feedback, consistent with control-theoretic analysis.[125] This work demonstrates the promise of formal modeling and feedback principles to engineer safer and more reliable cell-based therapies. At the same time, the broader challenge remains translating such designs from theory into reliable biological parts while capturing biophysical constraints. Moving forward, close collaboration between control theorists and genetic engineers will be essential to advance control-informed CAR T design.[126]

In clinical CAR T cell therapy, control theory has guided clinical dosing strategies. For example, in aggressive brain tumor glioblastoma, where prolonging survival is the primary goal, Bodzioch et al. applied optimal control theory (a mathematical framework that determines optimal control inputs by balancing the best possible outcome with system constraints) to design dosing schedules that minimize tumor burden, limit cumulative CAR T dosing, and reduce the duration of time the tumor remains near critical levels.[127] The problem was solved analytically and numerically to determine switching times in bang–bang control schedules for dosing. The results align with clinical observations of periodic CAR T administration in glioblastoma patients and suggest practicality, particularly when combined with complementary strategies such as checkpoint inhibition (which blocks inhibitory immune pathways) and multi-antigen targeting (which programs CAR T cells to recognize multiple tumor markers), which together help overcome both immunosuppression and tumor heterogeneity.[7,128] A similar approach could be applied to other clinical models, such as Owens et al.'s work on patient preconditioning regimens, where optimal control theory could help define theoretical best-case lymphodepletion dosing strategies to improve tumor control.

Moving forward, realizing this vision will depend on transforming control theory into a unifying design paradigm for CAR T interventions at both the molecular and clinical scales, made possible through the collaboration of theorists, engineers, synthetic biologists, and clinicians.

## CONCLUSIONS

The field of CAR T cell engineering has seen extraordinary acceleration and development in the last 8 years since the first FDA-approved therapy. Despite this success, its broader impact remains constrained by biological and clinical hurdles, ranging from adverse effects and antigen escape to limited efficacy in solid tumors and complex interpatient variability. As the therapeutic landscape diversifies—incorporating advanced CAR designs, new immune cell types, and combinational treatments—the need for rational, predictive, and integrative design frameworks becomes more urgent. In this perspective, we have outlined how mathematical and computational modeling can be systematically reframed and embedded within the translational research goals of CAR T cell therapy. Rather than treating modeling as a siloed academic discipline, we present it as a strategic toolset to tackle real-world therapeutic challenges: optimizing dosing, predicting patient-specific outcomes, managing adverse events, designing safer and more effective receptors, and expanding cell therapy to novel targets and cell types. We emphasize that no single modeling paradigm suffices to capture the full complexity of CAR T cell therapy. Instead, integrating complementary approaches—ODE-based modeling, agent-based simulations, control theory, multiscale hybrid frameworks, and data-driven machine learning—can align modeling efforts with specific clinical objectives, overall improving CAR T efficacy.

**BOX 1: GLOSSARY OF KEY TERMS AND ABBREVIATIONS**

**Biological Terms**

- **CAR**: Chimeric antigen receptor, an engineered receptor enabling T cells to recognize target antigens
- **IL**: Interleukin, a family of cytokines that mediate immune signaling
- **TGF-β:** Transforming growth factor beta, a clinically relevant immunosuppressive cytokine
- **TME**: Tumor microenvironment, the cellular and molecular milieu surrounding tumors
- **CD19**: Cluster of differentiation 19, a B-cell surface marker frequently targeted in CAR T therapies
- **EGFRvIII**: Epidermal growth factor receptor variant III, an antigen target for glioblastoma
- **HER2**: Human epidermal growth factor receptor 2, a cell surface protein commonly overexpressed in many solid tumors
- **CRS**: Cytokine release syndrome, systemic inflammatory response triggered by CAR T activation
- **ICANS**: Immune effector cell–associated neurotoxicity syndrome, a neurological side effect of CAR T therapy
- **MΦ**: Macrophage, innate immune cell involved in antigen presentation and cytokine secretion
- **NK**: Natural killer cell, innate lymphocyte with cytotoxic activity against tumors

**Modeling Terms**

- **PK**: Pharmacokinetics, modeling drug concentration over time
- **PD**: Pharmacodynamics, modeling drug effect on biological systems
- **ML**: Machine learning, computational methods for data-driven model discovery
- **ODE**: Ordinary differential equation, models temporal dynamics of continuous systems
- **PDE**: Partial differential equation, models spatiotemporal dynamics
- **ABM**: Agent-based model, simulates discrete cell–cell interactions

**Drug Names**

- **FDA**: U.S. Food and Drug Administration, regulatory agency approving therapies
- **Axi-cel**: Axicabtagene ciloleucel, FDA-approved CD19-directed CAR T cell therapy
- **Tisa-cel**: Tisagenlecleucel, FDA-approved CD19-directed CAR T cell therapy

**Cancer Types**

- **B-ALL**: B-cell acute lymphoblastic leukemia
- **CLL**: Chronic lymphocytic leukemia
- **DLBCL**: Diffuse large B-cell lymphoma
- **PMBCL**: Primary mediastinal B-cell lymphoma
- **MM**: Multiple myeloma
- **GBM**: Glioblastoma multiforme
- **mCRPC**: Metastatic castration-resistant prostate cancer

**Table 1: Challenges in CAR T Cell Therapy**

| Challenge | Description | Incidence | Proposed Solutions |
|---|---|---|---|
| **1. Cytokine Release Syndrome (CRS)** | Systemic inflammatory response caused by CAR T activation; characterized by fever, hypotension, and organ dysfunction.[129] | • Overall: Up to 80% (42–100%)[130]<br>• Severe CRS: 0–46%[131]<br>• Other CRS: 54–91%[132] | • Tocilizumab (anti-IL-6 receptor antibody)<br>• Corticosteroids (anti-inflammatory drugs)[130] |
| **2. Immune effector Cell-Associated Neurotoxicity Syndrome (ICANS)** | CAR T-induced neurotoxicity with symptoms from confusion to seizures and coma.[133,134] | • Overall: 20–70%<br>• Severe: ~56% of lymphoma cases[133,134] | • Symptom management<br>• Corticosteroids<br>• Close monitoring during treatment[130] |
| **3. Antigen Escape** | Tumor cells lose or downregulate target antigen (e.g., CD19), enabling immune evasion and relapse.[135] | • 20–28% in B-cell lymphoma patients<br>• 16–68% in B-cell acute lymphoblastic leukemia (B-ALL) patients[136] | • Tandem CAR T cells (recognize multiple targets with same CAR)<br>• Multi-CAR T cells (bicistronic designs)[135] |
| **4. Poor Infiltration into Solid Tumors** | Solid tumors present physical (extracellular matrix, vasculature) and immunosuppressive barriers to CAR T access and function.[137–139] | • Major limitation in treating non-hematologic cancers.[140]<br>• The hostile tumor microenvironment (TME) in solid tumors severely prevents CAR T cells from migrating, infiltrating, and killing. [141] | • “Armored” CAR T cells<br>• Genetically enhancing CAR T cell infiltration<br>• Neutralizing immunosuppressive cells[129]<br>• Modulating the TME for improved CAR-T infiltration. [141] |
| **5. On-Target, Off-Tumor Toxicity** | CARs recognize antigens also expressed on healthy tissues, causing unintended damage.[142] | • All CD19-directed CAR T cells can cause B-cell aplasia (loss of normal B cells)[143,144]<br>• HER2-directed CAR T cells have caused fatal toxicity due to non-specificity.[67] | • Logic-gated CARs that require multiple signals for activation<br>• Antigen-density sensing CAR T cells<br>• Immunoglobulin replacement for B-cell aplasia[129] |
| **6. Limited Persistence & T cell Exhaustion** | CAR T cells lose proliferative and killing capacity due to chronic antigen exposure and inhibitory signals in TME.[145] | • Major contributor to relapse and treatment failure in solid tumors.[146,147] | • Memory-like designs<br>• Checkpoint blockade (remove inhibitory signals)<br>• Metabolic reprogramming (improve energy use and sustain activity)[148,149] |

**Table 2: Challenges in CAR T Cell Modeling**

| Modeling Challenge | Biological basis | Impact on CAR T function | Implications for modeling |
|---|---|---|---|
| **1. Changing Cell States** | CAR T cells are dynamic and can alter their functional state in response to their environment.[150] Common transitions include activation, exhaustion, or differentiation into regulatory phenotypes. | T cell exhaustion due to chronic antigen exposure or immunosuppressive signaling reduces cytotoxicity, proliferation, and persistence.[152,153] | Models must account for time-dependent state transitions, often governed by stochastic or nonlinear dynamics, making them complex to predict and calibrate.[45,154] |
| **2.Interactions with other Immune Cells** | CAR T cells interact with many cell types in the TME, such as Tregs, MDSCs, and TAMs, which can inhibit CAR T activity via cytokines or cell-contact mechanisms.[155,156] | Immunosuppressive cells suppress CAR T function, reduce infiltration, and promote exhaustion.[157] Pro-inflammatory cells can enhance activity.[158] | Modeling must capture intercellular communication networks, requiring multi-agent or multi-compartment frameworks. |
| **3. Spatial Dynamics in Solid Tumors** | Solid tumors have dense extracellular matrix, irregular vasculature, and spatially heterogeneous antigen expression.[33,129] These act as physical and chemical barriers to CAR T infiltration and function.[159] | CAR T cells may fail to penetrate tumor cores or encounter antigen-negative zones, leading to relapse.[135] | Requires spatially resolved modeling rather than well-mixed assumptions. |

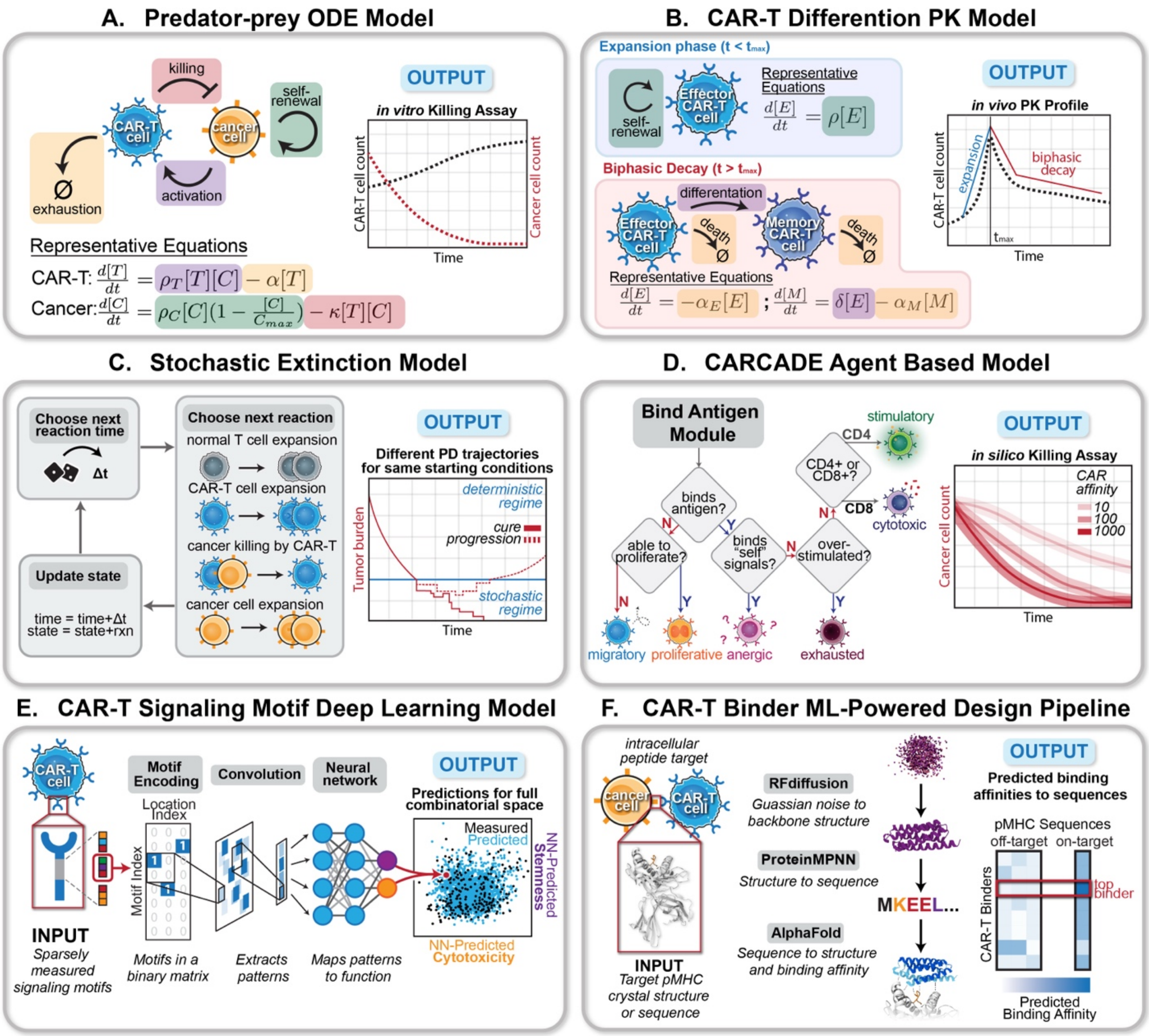

**Figure 1:** Various models of CAR T cell therapy employing diverse mathematical tools. A) ODE-based predator prey model of CAR T and cancer cell population growth dynamics.[12] B) Mechanistic PK model of CAR T population dynamics *in vivo*. CAR T cells differentiate into long-lived memory cells to explain the observed biphasic decay of CAR T cells.[18] C) Stochastic extinction model which uses the Gillespie algorithm to generate single trajectories for the chemical master equation, which can be mathematically related to differential equation-based PD dynamics of CAR T cells.[21] D) Hybrid PDE and agent-based model of CAR T cell and cancer cell populations. Depicted is a modified version of the a small portion of the decision tree that CAR T cells use to change cells state.[24] E) Simplified representation of the architecture of a deep learning model that can relate CAR T costimulatory domain combinations to CAR T cytotoxicity and stemness.[28] F) Simplified representation of a ML-powered design pipeline that can be used to generate *de novo* CAR T binders to pMHC structures on cancer cells.[29–31]

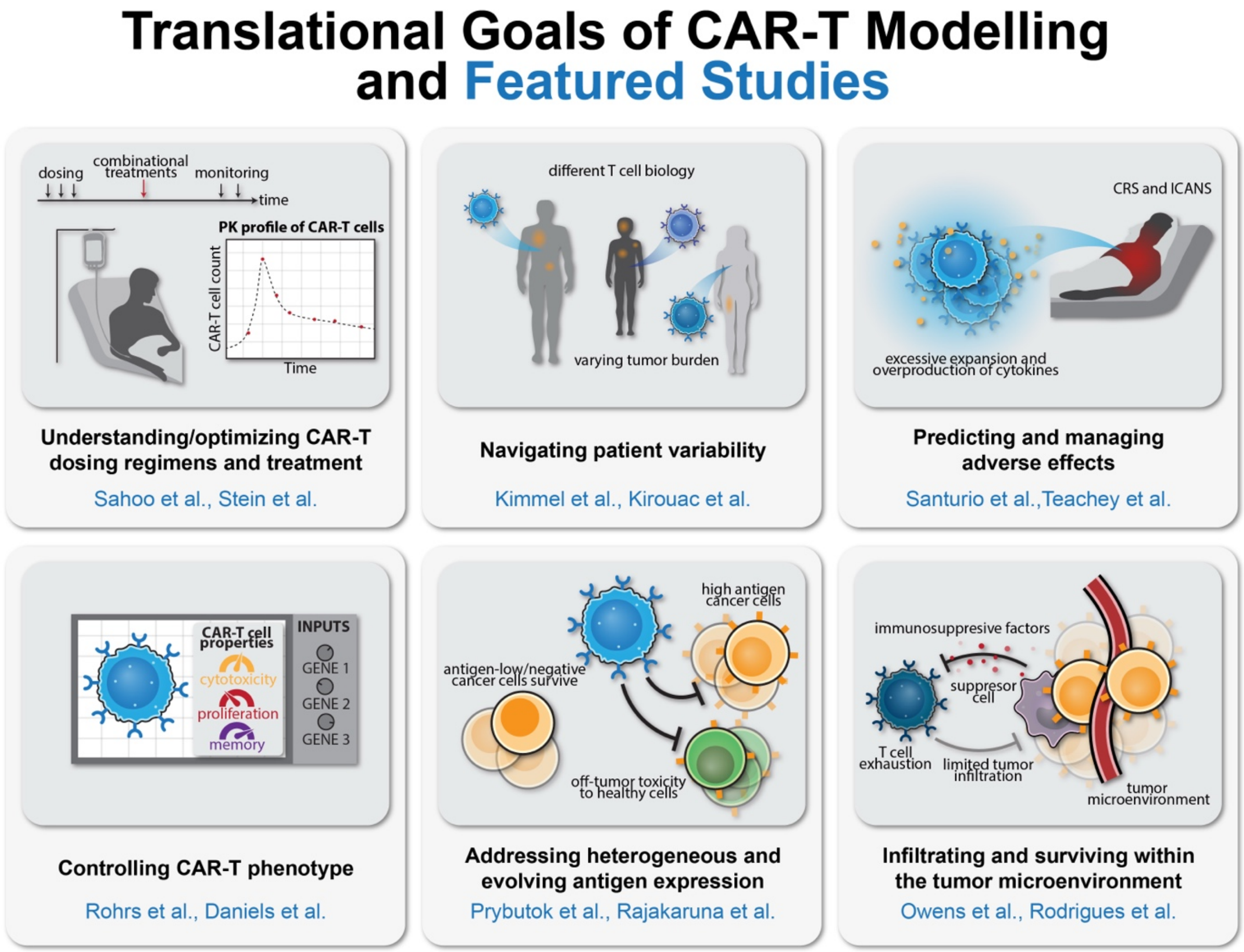

**Figure 2:** Translational goals of CAR T modeling and featured studies.

**BOX 2: DETERMINING THE MOST APPORPRIATE MODEL**

Choosing the appropriate mathematical model to represent CAR T cell dynamics is a critical step in extracting meaningful biological insights and guiding therapeutic optimization. Traditionally, mechanistic surrogate models are constructed from first principles based on known or hypothesized biological interactions derived from literature. While mechanistically grounded, such models are inherently hypothesis-driven, making them vulnerable to author bias and limited by the assumptions baked into their formulation. One approach to deal with this is to design several different systems of equations based on different theories about the system dynamics and use an objective measurement for the goodness-of-fit to determine the most accurate measure, such as information criteria, two of which are outlined in the table below.

| Model | Akaike Information Criterion (AIC) | Bayesian Information Criterion (BIC) |
|---|---|---|
| **Formula** | $2k - 2\ln(L)$ | $k\ln(n) - 2\ln(L)$ |
| **Strength** | Prediction accuracy | Choosing the correct model |

$L$ = maximized likelihood of the model
$k$ = number of estimated parameters
$n$ = number of observations

Both criteria are based around rewarding models with good fit (large $L$) and penalizing models with many parameters (large $k$). Although similar, AIC is most appropriate for choosing the model that will be most accurate at predicting future observations, while the BIC is more appropriate for finding the correct model to describe the system.[160] These criteria have been applied to models of CAR T cell dynamics.[48]

However, another approach to tackle this problem is to use data-driven techniques to determine the most appropriate system dynamics out of a set of candidate dynamics. Brunton et al. developed an algorithm called Sparse Identification of Nonlinear Dynamics (SINDy)[161] that was applied to CAR T cell system dynamics discovery by Brummer et al.[162] The SINDy algorithm generates sparse differential equations that capture the dynamics of the data with as few terms as possible. This is done by creating a library of terms that are functions of the data that could be used to fit the system, such as polynomials or other relevant functions like Michaelis-Menten terms or Hill functions for biological systems. Once the library is defined, best fit models are generated by minimizing the following equation,

$$||\Theta(x)\Xi - \dot{x}||_2 + \lambda||\Xi||_1$$

where $\Xi$ is the vector of coefficients for every function in the library and $\lambda$ is the regularization parameter. The first part of the loss function is the least squares fit while the second is the lasso which promotes sparsity in $\Xi$. As $\lambda$ is increased, the solution prioritizes 0s in $\Xi$, reducing the number of terms and decreasing the complexity of the model.

When applied to CAR T cell system dynamics, Brummer et al. included several starter functions based on previous ODE models of CAR T cells such as those in Sahoo et al.'s CARRGO model and following papers.[12,35] They applied SINDy to determine the best fit models, and discovered that different effector-to-target ratios were best described by different governing dynamics, such as logistic growth vs. weak allee affect growth of cancer cells.

The selection of modeling approach, whether based on information criteria, sparse regression methods like SINDy, or machine learning techniques should be guided by the nature and quantity of available data. Each framework offers distinct strengths, and when applied appropriately, can yield interpretable models that balance complexity, predictive power, and biological relevance in advancing CAR T therapy design.

**Table 3: Pros and cons of CAR T, CAR MΦ, CAR NK**

| Therapy | Pros | Cons |
|---|---|---|
| **CAR T** | • Highly cytotoxic<br>• Highly proliferative<br>• Memory phenotype can persist *in vivo* for decades | • Currently must be patient-derived<br>• Safety concerns<br>• Limited solid tumor infiltration<br>• Exhaustion / immunosuppression<br>• Antigen-negative relapse |
| **CAR M** | • Immune-modulatory<br>• Phagocytosis / antigen-presentation<br>• Natural solid tumor infiltration | • Currently must be patient-derived<br>• Not proliferative<br>• Less efficient killing dynamics<br>• No long-lived memory phenotype |
| **CAR NK** | • Highly cytotoxic<br>• Potential for donor- and cell line- derived therapy<br>• Antigen dependent and independent killing<br>• Superior safety profile | • Limited persistence *in vivo*<br>• Limited tumor infiltration<br>• Exhaustion / immunosuppression |

## CITATIONS

1. Lim, W.A., and June, C.H. (2017). The Principles of Engineering Immune Cells to Treat Cancer. Cell *168*, 724–740. https://doi.org/10.1016/j.cell.2017.01.016.

2. Mitra, A., Barua, A., Huang, L., Ganguly, S., Feng, Q., and He, B. (2023). From bench to bedside: the history and progress of CAR T cell therapy. Front. Immunol. *14*, 1188049. https://doi.org/10.3389/fimmu.2023.1188049.

3. Asmamaw Dejenie, T., Tiruneh G/Medhin, M., Dessie Terefe, G., Tadele Admasu, F., Wale Tesega, W., and Chekol Abebe, E. (2022). Current updates on generations, approvals, and clinical trials of CAR T-cell therapy. Hum. Vaccines Immunother. *18*, 2114254. https://doi.org/10.1080/21645515.2022.2114254.

4. Zheng, Z., Li, S., Liu, M., Chen, C., Zhang, L., and Zhou, D. (2023). Fine-Tuning through Generations: Advances in Structure and Production of CAR-T Therapy. Cancers *15*, 3476. https://doi.org/10.3390/cancers15133476.

5. Cao, J.-X., Gao, W.-J., You, J., Wu, L.-H., Liu, J.-L., and Wang, Z.-X. (2019). The efficacy of anti-CD19 chimeric antigen receptor T cells for B-cell malignancies. Cytotherapy *21*, 769–781. https://doi.org/10.1016/j.jcyt.2019.04.005.

6. Goyco Vera, D., Waghela, H., Nuh, M., Pan, J., and Lulla, P. (2024). Approved CAR-T therapies have reproducible efficacy and safety in clinical practice. Hum. Vaccines Immunother. *20*, 2378543. https://doi.org/10.1080/21645515.2024.2378543.

7. Rafiq, S., Hackett, C.S., and Brentjens, R.J. (2020). Engineering strategies to overcome the current roadblocks in CAR T cell therapy. Nat. Rev. Clin. Oncol. *17*, 147–167. https://doi.org/10.1038/s41571-019-0297-y.

8. Srivastava, S., and Riddell, S.R. (2015). Engineering CAR-T cells: Design concepts. Trends Immunol. *36*, 494–502. https://doi.org/10.1016/j.it.2015.06.004.

9. Lesko, L.J. (2021). Perspective on model-informed drug development. CPT Pharmacomet. Syst. Pharmacol. *10*, 1127–1129. https://doi.org/10.1002/psp4.12699.

10. Santurio, D.S., Barros, L.R.C., Glauche, I., and Fassoni, A.C. (2025). Mathematical modeling unveils the timeline of CAR-T cell therapy and macrophage-mediated cytokine release syndrome. PLOS Comput. Biol. *21*, e1012908. https://doi.org/10.1371/journal.pcbi.1012908.

11. Murias-Closas, A., Prats, C., Calvo, G., López-Codina, D., and Olesti, E. (2025). Computational modelling of CAR T-cell therapy: from cellular kinetics to patient-level predictions. eBioMedicine *113*, 105597. https://doi.org/10.1016/j.ebiom.2025.105597.

12. Sahoo, P., Yang, X., Abler, D., Maestrini, D., Adhikarla, V., Frankhouser, D., Cho, H., Machuca, V., Wang, D., Barish, M., et al. (2020). Mathematical deconvolution of CAR T-cell proliferation and exhaustion from real-time killing assay data. J. R. Soc. Interface *17*, 20190734. https://doi.org/10.1098/rsif.2019.0734.

13. Barros, L.R.C., Paixão, E.A., Valli, A.M.P., Naozuka, G.T., Fassoni, A.C., and Almeida, R.C. (2021). CARTmath—A Mathematical Model of CAR-T Immunotherapy in Preclinical Studies of Hematological Cancers. Cancers *13*, 2941. https://doi.org/10.3390/cancers13122941.

14. León-Triana, O., Pérez-Martínez, A., Ramírez-Orellana, M., and Pérez-García, V.M. (2021). Dual-Target CAR-Ts with On- and Off-Tumour Activity May Override Immune Suppression in Solid Cancers: A Mathematical Proof of Concept. Cancers *13*, 703. https://doi.org/10.3390/cancers13040703.

15. León-Triana, O., Sabir, S., Calvo, G.F., Belmonte-Beitia, J., Chulián, S., Martínez-Rubio, Á., Rosa, M., Pérez-Martínez, A., Ramirez-Orellana, M., and Pérez-García, V.M. (2021). CAR T cell therapy in B-cell acute lymphoblastic leukaemia: Insights from mathematical models. Commun. Nonlinear Sci. Numer. Simul. *94*, 105570. https://doi.org/10.1016/j.cnsns.2020.105570.

16. Santurio, D.S., Paixão, E.A., Barros, L.R.C., Almeida, R.C., and Fassoni, A.C. (2024). Mechanisms of resistance to CAR-T cell immunotherapy: Insights from a mathematical model. Appl. Math. Model. *125*, 1–15. https://doi.org/10.1016/j.apm.2023.08.029.

17. Putignano, G., Ruipérez-Campillo, S., Yuan, Z., Millet, J., and Guerrero-Aspizua, S. (2025). Mathematical models and computational approaches in CAR-T therapeutics. Front. Immunol. *16*, 1581210. https://doi.org/10.3389/fimmu.2025.1581210.

18. Stein, A.M., Grupp, S.A., Levine, J.E., Laetsch, T.W., Pulsipher, M.A., Boyer, M.W., August, K.J., Levine, B.L., Tomassian, L., Shah, S., et al. (2019). Tisagenlecleucel Model-Based Cellular Kinetic Analysis of Chimeric Antigen Receptor–T Cells. CPT Pharmacomet. Syst. Pharmacol. *8*, 285–295. https://doi.org/10.1002/psp4.12388.

19. Singh, A.P., Chen, W., Zheng, X., Mody, H., Carpenter, T.J., Zong, A., and Heald, D.L. (2021). Bench-to-bedside translation of chimeric antigen receptor (CAR) T cells using a multiscale systems pharmacokinetic-pharmacodynamic model: A case study with anti-BCMA CAR-T. CPT Pharmacomet. Syst. Pharmacol. *10*, 362–376. https://doi.org/10.1002/psp4.12598.

20. Zhou, T., Yang, J., Tan, Y., and Liu, Z. (2025). Threshold dynamics of a stochastic tumor-immune model combined oncolytic virus and chimeric antigen receptor T cell therapies. Chaos Solitons Fractals *197*, 116418. https://doi.org/10.1016/j.chaos.2025.116418.

21. Kimmel, G.J., Locke, F.L., and Altrock, P.M. (2021). The roles of T cell competition and stochastic extinction events in chimeric antigen receptor T cell therapy. Proc. R. Soc. B Biol. Sci. *288*, rspb.2021.0229, 20210229. https://doi.org/10.1098/rspb.2021.0229.

22. Shah, V., Womack, J., Zamora, A.E., Terhune, S.S., and Dash, R.K. (2023). Simulating the Evolution of Signaling Signatures During CART-Cell and Tumor Cell Interactions. In 2023 45th Annual International Conference of the IEEE Engineering in Medicine & Biology Society (EMBC) (IEEE), pp. 1–5. https://doi.org/10.1109/embc40787.2023.10340076.

23. Woelke, A.L., Murgueitio, M.S., and Preissner, R. (2010). Theoretical modeling techniques and their impact on tumor immunology. Clin. Dev. Immunol. *2010*, 271794. https://doi.org/10.1155/2010/271794.

24. Prybutok, A.N., Yu, J.S., Leonard, J.N., and Bagheri, N. (2022). Mapping CAR T-Cell Design Space Using Agent-Based Models. Front. Mol. Biosci. *9*, 849363. https://doi.org/10.3389/fmolb.2022.849363.

25. Rohrs, J.A., Zheng, D., Graham, N.A., Wang, P., and Finley, S.D. (2018). Computational Model of Chimeric Antigen Receptors Explains Site-Specific Phosphorylation Kinetics. Biophys. J. *115*, 1116–1129. https://doi.org/10.1016/j.bpj.2018.08.018.

26. Cess, C.G., and Finley, S.D. (2020). Data-driven analysis of a mechanistic model of CAR T cell signaling predicts effects of cell-to-cell heterogeneity. J. Theor. Biol. *489*, 110125. https://doi.org/10.1016/j.jtbi.2019.110125.

27. Harris, L.A., Hogg, J.S., Tapia, J.-J., Sekar, J.A.P., Gupta, S., Korsunsky, I., Arora, A., Barua, D., Sheehan, R.P., and Faeder, J.R. (2016). BioNetGen 2.2: advances in rule-based modeling. Bioinformatics *32*, 3366–3368. https://doi.org/10.1093/bioinformatics/btw469.

28. Daniels, K.G., Wang, S., Simic, M.S., Bhargava, H.K., Capponi, S., Tonai, Y., Yu, W., Bianco, S., and Lim, W.A. (2022). Decoding CAR T cell phenotype using combinatorial signaling motif libraries and machine learning. Science *378*, 1194–1200. https://doi.org/10.1126/science.abq0225.

29. Householder, K.D., Xiang, X., Jude, K.M., Deng, A., Obenaus, M., Zhao, Y., Wilson, S.C., Chen, X., Wang, N., and Garcia, K.C. (2025). De novo design and structure of a peptide–centric TCR mimic binding module. Science *389*, 375–379. https://doi.org/10.1126/science.adv3813.

30. Johansen, K.H., Wolff, D.S., Scapolo, B., Fernández-Quintero, M.L., Risager Christensen, C., Loeffler, J.R., Rivera-de-Torre, E., Overath, M.D., Kjærgaard Munk, K., Morell, O., et al. (2025). De novo-designed pMHC binders facilitate T cell–mediated cytotoxicity toward cancer cells. Science *389*, 380–385. https://doi.org/10.1126/science.adv0422.

31. Liu, B., Greenwood, N.F., Bonzanini, J.E., Motmaen, A., Meyerberg, J., Dao, T., Xiang, X., Ault, R., Sharp, J., Wang, C., et al. (2025). Design of high-specificity binders for peptide–MHC-I complexes. Science *389*, 386–391. https://doi.org/10.1126/science.adv0185.

32. Newick, K., Moon, E., and Albelda, S.M. (2016). Chimeric antigen receptor T-cell therapy for solid tumors. Mol. Ther. - Oncolytics *3*, 16006. https://doi.org/10.1038/mto.2016.6.

33. Hou, A.J., Chen, L.C., and Chen, Y.Y. (2021). Navigating CAR-T cells through the solid-tumour microenvironment. Nat. Rev. Drug Discov. *20*, 531–550. https://doi.org/10.1038/s41573-021-00189-2.

34. O'Rourke, D.M., Nasrallah, M.P., Desai, A., Melenhorst, J.J., Mansfield, K., Morrissette, J.J.D., Martinez-Lage, M., Brem, S., Maloney, E., Shen, A., et al. (2017). A single dose of peripherally infused EGFRvIII-directed CAR T cells mediates antigen loss and induces adaptive resistance in patients with recurrent glioblastoma. Sci. Transl. Med. *9*. https://doi.org/10.1126/scitranslmed.aaa0984.

35. Li, R., Sahoo, P., Wang, D., Wang, Q., Brown, C.E., Rockne, R.C., and Cho, H. (2023). Modeling interaction of Glioma cells and CAR T-cells considering multiple CAR T-cells bindings. ImmunoInformatics *9*, 100022. https://doi.org/10.1016/j.immuno.2023.100022.

36. Liu, C., Ayyar, V.S., Zheng, X., Chen, W., Zheng, S., Mody, H., Wang, W., Heald, D., Singh, A.P., and Cao, Y. (2021). Model-Based Cellular Kinetic Analysis of Chimeric Antigen Receptor-T Cells in Humans. Clin. Pharmacol. Ther. *109*, 716–727. https://doi.org/10.1002/cpt.2040.

37. Hardiansyah, D., and Ng, C.M. (2019). Quantitative Systems Pharmacology Model of Chimeric Antigen Receptor T-Cell Therapy. Clin. Transl. Sci. *12*, 343–349. https://doi.org/10.1111/cts.12636.

38. Martínez-Rubio, Á., Chulián, S., Blázquez Goñi, C., Ramírez Orellana, M., Pérez Martínez, A., Navarro-Zapata, A., Ferreras, C., Pérez-García, V.M., and Rosa, M. (2021). A Mathematical Description of the Bone Marrow Dynamics during CAR T-Cell Therapy in B-Cell Childhood Acute Lymphoblastic Leukemia. Int. J. Mol. Sci. *22*, 6371. https://doi.org/10.3390/ijms22126371.

39. Sinelshchikov, D., Belmonte-Beitia, J., and Italia, M. (2025). A mathematical model of CAR-T cell therapy in combination with chemotherapy for malignant gliomas. Chaos Interdiscip. J. Nonlinear Sci. *35*, 063104. https://doi.org/10.1063/5.0260252.

40. Owens, K., and Bozic, I. (2021). Modeling CAR T-Cell Therapy with Patient Preconditioning. Bull. Math. Biol. *83*, 42. https://doi.org/10.1007/s11538-021-00869-5.

41. Sgouros, G., Bodei, L., McDevitt, M.R., and Nedrow, J.R. (2020). Radiopharmaceutical therapy in cancer: clinical advances and challenges. Nat. Rev. Drug Discov. *19*, 589–608. https://doi.org/10.1038/s41573-020-0073-9.

42. St. James, S., Bednarz, B., Benedict, S., Buchsbaum, J.C., Dewaraja, Y., Frey, E., Hobbs, R., Grudzinski, J., Roncali, E., Sgouros, G., et al. (2021). Current Status of Radiopharmaceutical Therapy. Int. J. Radiat. Oncol. *109*, 891–901. https://doi.org/10.1016/j.ijrobp.2020.08.035.

43. Adhikarla, V., Awuah, D., Brummer, A.B., Caserta, E., Krishnan, A., Pichiorri, F., Minnix, M., Shively, J.E., Wong, J.Y.C., Wang, X., et al. (2021). A Mathematical Modeling Approach for Targeted Radionuclide and Chimeric Antigen Receptor T Cell Combination Therapy. Cancers *13*, 5171. https://doi.org/10.3390/cancers13205171.

44. Laetsch, T.W., Maude, S.L., Rives, S., Hiramatsu, H., Bittencourt, H., Bader, P., Baruchel, A., Boyer, M., De Moerloose, B., Qayed, M., et al. (2023). Three-Year Update of Tisagenlecleucel in Pediatric and Young Adult Patients With Relapsed/Refractory Acute Lymphoblastic Leukemia in the ELIANA Trial. J. Clin. Oncol. *41*, 1664–1669. https://doi.org/10.1200/jco.22.00642.

45. Hoang, C., Phan, T.A., Turtle, C.J., and Tian, J.P. (2024). A stochastic framework for evaluating CAR T cell therapy efficacy and variability. Math. Biosci. *368*, 109141. https://doi.org/10.1016/j.mbs.2024.109141.

46. Liu, L., Ma, C., Zhang, Z., Witkowski, M.T., Aifantis, I., Ghassemi, S., and Chen, W. (2022). Computational model of CAR T-cell immunotherapy dissects and predicts leukemia patient responses at remission, resistance, and relapse. J. Immunother. Cancer *10*, e005360. https://doi.org/10.1136/jitc-2022-005360.

47. Paixão, E.A., Barros, L.R.C., Fassoni, A.C., and Almeida, R.C. (2022). Modeling Patient-Specific CAR-T Cell Dynamics: Multiphasic Kinetics via Phenotypic Differentiation. Cancers *14*, 5576. https://doi.org/10.3390/cancers14225576.

48. Kirouac, D.C., Zmurchok, C., Deyati, A., Sicherman, J., Bond, C., and Zandstra, P.W. (2023). Deconvolution of clinical variance in CAR-T cell pharmacology and response. Nat. Biotechnol. *41*, 1606–1617. https://doi.org/10.1038/s41587-023-01687-x.

49. Finney, O.C., Brakke, H., Rawlings-Rhea, S., Hicks, R., Doolittle, D., Lopez, M., Futrell, B., Orentas, R.J., Li, D., Gardner, R., et al. CD19 CAR T cell product and disease attributes predict leukemia remission durability. J. Clin. Invest. *129*, 2123–2132. https://doi.org/10.1172/JCI125423.

50. Tong, Y., Udupa, J.K., Chong, E., Winchell, N., Sun, C., Zou, Y., Schuster, S.J., and Torigian, D.A. (2023). Prediction of lymphoma response to CAR T cells by deep learning-based image analysis. PLOS ONE *18*, e0282573. https://doi.org/10.1371/journal.pone.0282573.

51. Jak, M., van der Velden, B.H.M., de Keizer, B., Elias, S.G., Minnema, M.C., and Gilhuijs, K.G.A. (2022). Prediction of Poor Outcome after Tisagenlecleucel in Patients with Relapsed or Refractory Diffuse Large B Cell Lymphoma (DLBCL) Using Artificial Intelligence Analysis of Pre-Infusion PET/CT. Blood *140*, 1919–1920. https://doi.org/10.1182/blood-2022-158543.

52. Neelapu, S.S., Tummala, S., Kebriaei, P., Wierda, W., Gutierrez, C., Locke, F.L., Komanduri, K.V., Lin, Y., Jain, N., Daver, N., et al. (2018). Chimeric antigen receptor T-cell therapy — assessment and management of toxicities. Nat. Rev. Clin. Oncol. *15*, 47–62. https://doi.org/10.1038/nrclinonc.2017.148.

53. Maude, S.L., Frey, N., Shaw, P.A., Aplenc, R., Barrett, D.M., Bunin, N.J., Chew, A., Gonzalez, V.E., Zheng, Z., Lacey, S.F., et al. (2014). Chimeric Antigen Receptor T Cells for Sustained Remissions in Leukemia. N. Engl. J. Med. *371*, 1507–1517. https://doi.org/10.1056/NEJMoa1407222.

54. Neelapu, S.S., Locke, F.L., Bartlett, N.L., Lekakis, L.J., Miklos, D.B., Jacobson, C.A., Braunschweig, I., Oluwole, O.O., Siddiqi, T., Lin, Y., et al. (2017). Axicabtagene Ciloleucel CAR T-Cell Therapy in Refractory Large B-Cell Lymphoma. N. Engl. J. Med. *377*, 2531–2544. https://doi.org/10.1056/NEJMoa1707447.

55. Schuster, S.J., Bishop, M.R., Tam, C.S., Waller, E.K., Borchmann, P., McGuirk, J.P., Jäger, U., Jaglowski, S., Andreadis, C., Westin, J.R., et al. (2019). Tisagenlecleucel in Adult Relapsed or Refractory Diffuse Large B-Cell Lymphoma. N. Engl. J. Med. *380*, 45–56. https://doi.org/10.1056/NEJMoa1804980.

56. Teachey, D.T., Lacey, S.F., Shaw, P.A., Melenhorst, J.J., Maude, S.L., Frey, N., Pequignot, E., Gonzalez, V.E., Chen, F., Finklestein, J., et al. (2016). Identification of Predictive Biomarkers for Cytokine Release Syndrome after Chimeric Antigen Receptor T-cell Therapy for Acute Lymphoblastic Leukemia. Cancer Discov. *6*, 664–679. https://doi.org/10.1158/2159-8290.CD-16-0040.

57. Ferrer Lores, B., Serrano, A., Hernani, R., Sopena-Novales, P., Ventura Lopez, L., Teruel, A.I., Ortiz Algarra, A., Saus Carreres, A., Arroyo Martin, I., Vasile Tudorache, A.R., et al. (2023). CART-AI-Radiomics: Survival and Neurotoxicity Prediction in B-Cell Lymphoma Patients Treated with CAR-T Cells through an Imaging Features-Based Model. Blood *142*, 5017. https://doi.org/10.1182/blood-2023-185088.

58. Good, Z., Spiegel, J.Y., Sahaf, B., Malipatlolla, M.B., Ehlinger, Z.J., Kurra, S., Desai, M.H., Reynolds, W.D., Lin, A.W., Vandris, P., et al. (2022). Post-infusion CAR TReg cells identify patients resistant to CD19-CAR therapy. Nat. Med. *28*, 1860–1871. https://doi.org/10.1038/s41591-022-01960-7.

59. Harris, D.T., Hager, M.V., Smith, S.N., Cai, Q., Stone, J.D., Kruger, P., Lever, M., Dushek, O., Schmitt, T.M., Greenberg, P.D., et al. (2018). Comparison of T Cell Activities Mediated by Human TCRs and CARs That Use the Same Recognition Domains. J. Immunol. Baltim. Md 1950 *200*, 1088–1100. https://doi.org/10.4049/jimmunol.1700236.

60. Rohrs, J.A., Siegler, E.L., Wang, P., and Finley, S.D. (2020). ERK Activation in CAR T Cells Is Amplified by CD28-Mediated Increase in CD3ζ Phosphorylation. iScience *23*, 101023. https://doi.org/10.1016/j.isci.2020.101023.

61. Tserunyan, V., and Finley, S.D. (2022). Computational analysis of 4-1BB-induced NFκB signaling suggests improvements to CAR cell design. Cell Commun. Signal. CCS *20*, 129. https://doi.org/10.1186/s12964-022-00937-w.

62. Tserunyan, V., and Finley, S. (2024). Information-Theoretic Analysis of a Model of CAR-4-1BB-Mediated NFκB Activation. Bull. Math. Biol. *86*, 5. https://doi.org/10.1007/s11538-023-01232-6.

63. Bolouri, H., Young, M., Beilke, J., Johnson, R., Fox, B., Huang, L., Santini, C.C., Hill, C.M., Vries, A.-R.V.D.V.D., Shannon, P.T., et al. (2020). Integrative network modeling reveals mechanisms underlying T cell exhaustion. Sci. Rep. *10*, 1915. https://doi.org/10.1038/s41598-020-58600-8.

64. Hou, Y., Zak, J., Shi, Y., Pratumchai, I., Dinner, B., Wang, W., Qin, K., Weber, E.W., Teijaro, J.R., and Wu, P. (2025). Transient EZH2 Suppression by Tazemetostat during In Vitro Expansion Maintains T-Cell Stemness and Improves Adoptive T-Cell Therapy. Cancer Immunol. Res. *13*, 47–65. https://doi.org/10.1158/2326-6066.CIR-24-0089.

65. Chen, J., Qiu, S., Li, W., Wang, K., Zhang, Y., Yang, H., Liu, B., Li, G., Li, L., Chen, M., et al. (2023). Tuning charge density of chimeric antigen receptor optimizes tonic signaling and CAR-T cell fitness. Cell Res. *33*, 341–354. https://doi.org/10.1038/s41422-023-00789-0.

66. Qiu, S., Chen, J., Wu, T., Li, L., Wang, G., Wu, H., Song, X., Liu, X., and Wang, H. (2024). CAR-Toner: an AI-driven approach for CAR tonic signaling prediction and optimization. Cell Res. *34*, 386–388. https://doi.org/10.1038/s41422-024-00936-1.

67. Morgan, R.A., Yang, J.C., Kitano, M., Dudley, M.E., Laurencot, C.M., and Rosenberg, S.A. (2010). Case Report of a Serious Adverse Event Following the Administration of T Cells Transduced With a Chimeric Antigen Receptor Recognizing ERBB2. Mol. Ther. *18*, 843–851. https://doi.org/10.1038/mt.2010.24.

68. Giorgadze, T., Fischel, H., Tessier, A., and Norton, K.-A. (2022). Investigating Two Modes of Cancer-Associated Antigen Heterogeneity in an Agent-Based Model of Chimeric Antigen Receptor T-Cell Therapy. Cells *11*, 3165. https://doi.org/10.3390/cells11193165.

69. Santurio, D.S., and Barros, L.R.C. (2022). A Mathematical Model for On-Target Off-Tumor Effect of CAR-T Cells on Gliomas. Front. Syst. Biol. *2*. https://doi.org/10.3389/fsysb.2022.923085.

70. Kara, E., Jackson, T.L., Jones, C., Sison, R., and McGee Ii, R.L. (2024). Mathematical modeling insights into improving CAR T cell therapy for solid tumors with bystander effects. Npj Syst. Biol. Appl. *10*, 105. https://doi.org/10.1038/s41540-024-00435-4.

71. Jumper, J., Evans, R., Pritzel, A., Green, T., Figurnov, M., Ronneberger, O., Tunyasuvunakool, K., Bates, R., Žídek, A., Potapenko, A., et al. (2021). Highly accurate protein structure prediction with AlphaFold. Nature *596*, 583–589. https://doi.org/10.1038/s41586-021-03819-2.

72. Abramson, J., Adler, J., Dunger, J., Evans, R., Green, T., Pritzel, A., Ronneberger, O., Willmore, L., Ballard, A.J., Bambrick, J., et al. (2024). Accurate structure prediction of biomolecular interactions with AlphaFold 3. Nature *630*, 493–500. https://doi.org/10.1038/s41586-024-07487-w.

73. Dauparas, J., Anishchenko, I., Bennett, N., Bai, H., Ragotte, R.J., Milles, L.F., Wicky, B.I.M., Courbet, A., de Haas, R.J., Bethel, N., et al. (2022). Robust deep learning–based protein sequence design using ProteinMPNN. Science *378*, 49–56. https://doi.org/10.1126/science.add2187.

74. Watson, J.L., Juergens, D., Bennett, N.R., Trippe, B.L., Yim, J., Eisenach, H.E., Ahern, W., Borst, A.J., Ragotte, R.J., Milles, L.F., et al. (2023). De novo design of protein structure and function with RFdiffusion. Nature *620*, 1089–1100. https://doi.org/10.1038/s41586-023-06415-8.

75. Rajakaruna, H., Desai, M., and Das, J. (2023). PASCAR: a multiscale framework to explore the design space of constitutive and inducible CAR T cells. Life Sci. Alliance *6*. https://doi.org/10.26508/lsa.202302171.

76. McKeithan, T.W. (1995). Kinetic proofreading in T-cell receptor signal transduction. Proc. Natl. Acad. Sci. *92*, 5042–5046. https://doi.org/10.1073/pnas.92.11.5042.

77. Hopfield, J.J. (1974). Kinetic Proofreading: A New Mechanism for Reducing Errors in Biosynthetic Processes Requiring High Specificity. Proc. Natl. Acad. Sci. U. S. A. *71*, 4135–4139. https://doi.org/10.1073/pnas.71.10.4135.

78. Narayan, V., Barber-Rotenberg, J.S., Jung, I.-Y., Lacey, S.F., Rech, A.J., Davis, M.M., Hwang, W.-T., Lal, P., Carpenter, E.L., Maude, S.L., et al. (2022). PSMA-targeting

TGFβ-insensitive armored CAR T cells in metastatic castration-resistant prostate cancer: a phase 1 trial. Nat. Med. *28*, 724–734. https://doi.org/10.1038/s41591-022-01726-1.

79. Owens, K., Rahman, A., and Bozic, I. (2025). Spatiotemporal dynamics of tumor–CAR T-cell interaction following local administration in solid cancers. PLOS Comput. Biol. *21*, e1013117. https://doi.org/10.1371/journal.pcbi.1013117.

80. Bordel-Vozmediano, S., Sabir, S., Benito-Barca, L., Weigelin, B., and Pérez-García, V.M. (2025). Geometric immunosuppression in CAR T-cell treatment: Insights from mathematical modeling. Comput. Biol. Med. *194*, 110427. https://doi.org/10.1016/j.compbiomed.2025.110427.

81. Rodrigues, G., Silva, J.G. da, Santurio, D.S., and Mancera, P.F.A. (2024). A Mathematical Model to Describe the Melanoma Dynamics Under Effects of Macrophage Inhibition and CAR T-cell Therapy. Trends Comput. Appl. Math. *25*, e01734–e01734. https://doi.org/10.5540/tcam.2024.025.e01734.

82. Gong, C., Milberg, O., Wang, B., Vicini, P., Narwal, R., Roskos, L., and Popel, A.S. (2017). A computational multiscale agent-based model for simulating spatio-temporal tumour immune response to PD1 and PDL1 inhibition. J. R. Soc. Interface *14*, 20170320. https://doi.org/10.1098/rsif.2017.0320.

83. Mahlbacher, G., Curtis, L.T., Lowengrub, J., and Frieboes, H.B. (2018). Mathematical modeling of tumor-associated macrophage interactions with the cancer microenvironment. J. Immunother. Cancer *6*. https://doi.org/10.1186/s40425-017-0313-7.

84. Berman, H.M., Westbrook, J., Feng, Z., Gilliland, G., Bhat, T.N., Weissig, H., Shindyalov, I.N., and Bourne, P.E. (2000). The Protein Data Bank. Nucleic Acids Res. *28*, 235–242. https://doi.org/10.1093/nar/28.1.235.

85. Uhlén, M., Fagerberg, L., Hallström, B.M., Lindskog, C., Oksvold, P., Mardinoglu, A., Sivertsson, Å., Kampf, C., Sjöstedt, E., Asplund, A., et al. (2015). Tissue-based map of the human proteome. Science *347*, 1260419. https://doi.org/10.1126/science.1260419.

86. The UniProt Consortium (2023). UniProt: the Universal Protein Knowledgebase in 2023. Nucleic Acids Res. *51*, D523–D531. https://doi.org/10.1093/nar/gkac1052.

87. CZI Cell Science Program, Abdulla, S., Aevermann, B., Assis, P., Badajoz, S., Bell, S.M., Bezzi, E., Cakir, B., Chaffer, J., Chambers, S., et al. (2025). CZ CELLxGENE Discover: a single-cell data platform for scalable exploration, analysis and modeling of aggregated data. Nucleic Acids Res. *53*, D886–D900. https://doi.org/10.1093/nar/gkae1142.

88. Cao, L.-Y., Zhao, Y., Chen, Y., Ma, P., Xie, J.-C., Pan, X.-M., Zhang, X., Chen, Y.-C., Wang, Q., and Xie, L.-L. (2025). CAR-T cell therapy clinical trials: global progress, challenges, and future directions from ClinicalTrials.gov insights. Front. Immunol. *16*, 1583116. https://doi.org/10.3389/fimmu.2025.1583116.

89. Roohani, Y., Huang, K., and Leskovec, J. (2024). Predicting transcriptional outcomes of novel multigene perturbations with GEARS. Nat. Biotechnol. *42*, 927–935. https://doi.org/10.1038/s41587-023-01905-6.

90. Goodman, D.B., Azimi, C.S., Kearns, K., Talbot, A., Garakani, K., Garcia, J., Patel, N., Hwang, B., Lee, D., Park, E., et al. (2022). Pooled screening of CAR T cells identifies diverse immune signaling domains for next-generation immunotherapies. Sci. Transl. Med. *14*, eabm1463. https://doi.org/10.1126/scitranslmed.abm1463.

91. Dhainaut, M., Rose, S.A., Akturk, G., Wroblewska, A., Nielsen, S.R., Park, E.S., Buckup, M., Roudko, V., Pia, L., Sweeney, R., et al. (2022). Spatial CRISPR genomics identifies regulators of the tumor microenvironment. Cell *185*, 1223-1239.e20. https://doi.org/10.1016/j.cell.2022.02.015.

92. Binan, L., Jiang, A., Danquah, S.A., Valakh, V., Simonton, B., Bezney, J., Manguso, R.T., Yates, K.B., Nehme, R., Cleary, B., et al. (2025). Simultaneous CRISPR screening and spatial transcriptomics reveal intracellular, intercellular, and functional transcriptional circuits. Cell *188*, 2141-2158.e18. https://doi.org/10.1016/j.cell.2025.02.012.

93. Gong, Y., Fei, P., Zhang, Y., Xu, Y., and Wei, J. (2025). From Multi-Omics to Visualization and Beyond: Bridging Micro and Macro Insights in CAR-T Cell Therapy. Adv. Sci. *12*, 2501095. https://doi.org/10.1002/advs.202501095.

94. Mulazzani, M., Fräßle, S.P., Von Mücke-Heim, I., Langer, S., Zhou, X., Ishikawa-Ankerhold, H., Leube, J., Zhang, W., Dötsch, S., Svec, M., et al. (2019). Long-term in vivo microscopy of CAR T cell dynamics during eradication of CNS lymphoma in mice. Proc. Natl. Acad. Sci. *116*, 24275–24284. https://doi.org/10.1073/pnas.1903854116.

95. Cazaux, M., Grandjean, C.L., Lemaître, F., Garcia, Z., Beck, R.J., Milo, I., Postat, J., Beltman, J.B., Cheadle, E.J., and Bousso, P. (2019). Single-cell imaging of CAR T cell activity in vivo reveals extensive functional and anatomical heterogeneity. J. Exp. Med. *216*, 1038–1049. https://doi.org/10.1084/jem.20182375.

96. Liu, L., Yoon, C.W., Yuan, Z., Guo, T., Qu, Y., He, P., Yu, X., Zhu, Z., Limsakul, P., and Wang, Y. (2023). Cellular and molecular imaging of CAR-T cell-based immunotherapy. Adv. Drug Deliv. Rev. *203*, 115135. https://doi.org/10.1016/j.addr.2023.115135.

97. Cess, C.G., and Finley, S.D. (2023). Calibrating agent-based models to tumor images using representation learning. PLOS Comput. Biol. *19*, e1011070. https://doi.org/10.1371/journal.pcbi.1011070.

98. Lin, Z., Akin, H., Rao, R., Hie, B., Zhu, Z., Lu, W., Smetanin, N., Verkuil, R., Kabeli, O., Shmueli, Y., et al. (2023). Evolutionary-scale prediction of atomic-level protein structure with a language model. Science *379*, 1123–1130. https://doi.org/10.1126/science.ade2574.

99. Nguyen, E., Poli, M., Durrant, M.G., Kang, B., Katrekar, D., Li, D.B., Bartie, L.J., Thomas, A.W., King, S.H., Brixi, G., et al. (2024). Sequence modeling and design from molecular to genome scale with Evo. Science *386*, eado9336. https://doi.org/10.1126/science.ado9336.

100. Capponi, S., and Daniels, K.G. (2023). Harnessing the power of artificial intelligence to advance cell therapy. Immunol. Rev. *320*, 147–165. https://doi.org/10.1111/imr.13236.

101. Cess, C.G., and Finley, S.D. (2020). Multi-scale modeling of macrophage—T cell interactions within the tumor microenvironment. PLOS Comput. Biol. *16*, e1008519. https://doi.org/10.1371/journal.pcbi.1008519.

102. Tangella, N., Cess, C.G., Ildefonso, G.V., and Finley, S.D. (2024). Integrating mechanism-based T cell phenotypes into a model of tumor–immune cell interactions. APL Bioeng. *8*, 036111. https://doi.org/10.1063/5.0205996.

103. Namburi, S., Latif, T., Oluwole, O.O., Cross, S.J., Simmons, G., Iragavarapu, C., Hu, B., Jih, G., Bullaughey, K., Das, P.A., et al. (2024). Interim results from the ELiPSE-1 study: A phase 1, multicenter, open-label study of CNTY-101 in subjects with relapsed or refractory CD19-positive B-cell malignancies. J. Clin. Oncol. https://doi.org/10.1200/JCO.2024.42.16_suppl.7023.

104. Bachanova, V., Ghobadi, A., Patel, K., Park, J.H., Flinn, I.W., Shah, P., Wong, C., Bickers, C., Szabo, P., Wong, L., et al. (2021). Safety and Efficacy of FT596, a First-in-Class, Multi-Antigen Targeted, Off-the-Shelf, iPSC-Derived CD19 CAR NK Cell Therapy in Relapsed/Refractory B-Cell Lymphoma. Blood *138*, 823. https://doi.org/10.1182/blood-2021-151185.

105. Zhang, L., Meng, Y., Feng, X., and Han, Z. (2022). CAR-NK cells for cancer immunotherapy: from bench to bedside. Biomark. Res. *10*, 12. https://doi.org/10.1186/s40364-022-00364-6.

106. Reiss, K.A., Angelos, M.G., Dees, E.C., Yuan, Y., Ueno, N.T., Pohlmann, P.R., Johnson, M.L., Chao, J., Shestova, O., Serody, J.S., et al. (2025). CAR-macrophage therapy for HER2-overexpressing advanced solid tumors: a phase 1 trial. Nat. Med. *31*, 1171–1182. https://doi.org/10.1038/s41591-025-03495-z.

107. Marin, D., Li, Y., Basar, R., Rafei, H., Daher, M., Dou, J., Mohanty, V., Dede, M., Nieto, Y., Uprety, N., et al. (2024). Safety, efficacy and determinants of response of allogeneic CD19-specific CAR-NK cells in CD19+ B cell tumors: a phase 1/2 trial. Nat. Med. *30*, 772–784. https://doi.org/10.1038/s41591-023-02785-8.

108. Jørgensen, L.V., Christensen, E.B., Barnkob, M.B., and Barington, T. (2025). The clinical landscape of CAR NK cells. Exp. Hematol. Oncol. *14*, 46. https://doi.org/10.1186/s40164-025-00633-8.

109. Xu, S., and Liu, M. (2026). Mathematical model suggests current CAR-macrophage dosage is efficient to low pre-infusion tumour burden but refractory to high tumour burden. J. Theor. Biol. *616*, 112263. https://doi.org/10.1016/j.jtbi.2025.112263.

110. Look, T., Sankowski, R., Bouzereau, M., Fazio, S., Sun, M., Buck, A., Binder, N., Mastall, M., Prisco, F., Seehusen, F., et al. (2025). CAR T cells, CAR NK cells, and CAR macrophages exhibit distinct traits in glioma models but are similarly enhanced when combined with cytokines. Cell Rep. Med. *6*. https://doi.org/10.1016/j.xcrm.2025.101931.

111. Zhao, C., Mirando, A.C., Sové, R.J., Medeiros, T.X., Annex, B.H., and Popel, A.S. (2019). A mechanistic integrative computational model of macrophage polarization: Implications in human pathophysiology. PLoS Comput. Biol. *15*, e1007468. https://doi.org/10.1371/journal.pcbi.1007468.

112. Nickaeen, N., Ghaisari, J., Heiner, M., Moein, S., and Gheisari, Y. (2019). Agent-based modeling and bifurcation analysis reveal mechanisms of macrophage polarization and phenotype pattern distribution. Sci. Rep. *9*, 12764. https://doi.org/10.1038/s41598-019-48865-z.

113. Minucci, S.B., Heise, R.L., and Reynolds, A.M. (2024). Agent-based vs. equation-based multi-scale modeling for macrophage polarization. PLOS ONE *19*, e0270779. https://doi.org/10.1371/journal.pone.0270779.

114. Klichinsky, M., Ruella, M., Shestova, O., Lu, X.M., Best, A., Zeeman, M., Schmierer, M., Gabrusiewicz, K., Anderson, N.R., Petty, N.E., et al. (2020). Human chimeric antigen receptor macrophages for cancer immunotherapy. Nat. Biotechnol. *38*, 947–953. https://doi.org/10.1038/s41587-020-0462-y.

115. Lei, A., Yu, H., Lu, S., Lu, H., Ding, X., Tan, T., Zhang, H., Zhu, M., Tian, L., Wang, X., et al. (2024). A second-generation M1-polarized CAR macrophage with antitumor efficacy. Nat. Immunol. *25*, 102–116. https://doi.org/10.1038/s41590-023-01687-8.

116. Jørgensen, L.V., Christensen, E.B., Barnkob, M.B., and Barington, T. (2025). The clinical landscape of CAR NK cells. Exp. Hematol. Oncol. *14*, 46. https://doi.org/10.1186/s40164-025-00633-8.

117. Arabameri, A., and Arab, S. (2024). Understanding the Interplay of CAR-NK Cells and Triple-Negative Breast Cancer: Insights from Computational Modeling. Bull. Math. Biol. *86*, 20. https://doi.org/10.1007/s11538-023-01247-z.

118. Amoddeo, A. (2024). *In silico* assessment of CAR macrophages activity against SARS-CoV-2 infection. Heliyon *10*, e39689. https://doi.org/10.1016/j.heliyon.2024.e39689.

119. Liu, M., Liu, J., Liang, Z., Dai, K., Gan, J., Wang, Q., Xu, Y., Chen, Y.H., and Wan, X. (2022). CAR-Macrophages and CAR-T Cells Synergistically Kill Tumor Cells In Vitro. Cells *11*, 3692. https://doi.org/10.3390/cells11223692.

120. Fedorov, V.D., Themeli, M., and Sadelain, M. (2013). PD-1– and CTLA-4–Based Inhibitory Chimeric Antigen Receptors (iCARs) Divert Off-Target Immunotherapy Responses. Sci. Transl. Med. *5*. https://doi.org/10.1126/scitranslmed.3006597.

121. Guercio, M., Manni, S., Boffa, I., Caruso, S., Di Cecca, S., Sinibaldi, M., Abbaszadeh, Z., Camera, A., Ciccone, R., Polito, V.A., et al. (2021). Inclusion of the Inducible Caspase 9 Suicide Gene in CAR Construct Increases Safety of CAR.CD19 T Cell Therapy in B-Cell Malignancies. Front. Immunol. *12*, 755639. https://doi.org/10.3389/fimmu.2021.755639.

122. Weber, E.W., Parker, K.R., Sotillo, E., Lynn, R.C., Anbunathan, H., Lattin, J., Good, Z., Belk, J.A., Daniel, B., and Klysz, D. (2021). Transient rest restores functionality in exhausted CAR-T cells through epigenetic remodeling. Science *372*, eaba1786.

123. Hernandez-Lopez, R.A., Yu, W., Cabral, K.A., Creasey, O.A., Lopez Pazmino, M. del P., Tonai, Y., De Guzman, A., Mäkelä, A., Saksela, K., and Gartner, Z.J. (2021). T cell circuits that sense antigen density with an ultrasensitive threshold. Science *371*, 1166–1171.

124. Hart, Y., Reich-Zeliger, S., Antebi, Y.E., Zaretsky, I., Mayo, A.E., Alon, U., and Friedman, N. (2014). Paradoxical signaling by a secreted molecule leads to homeostasis of cell levels. Cell *158*, 1022–1032. https://doi.org/10.1016/j.cell.2014.07.033.

125. Ma, Y., Budde, M.W., Mayalu, M.N., Zhu, J., Lu, A.C., Murray, R.M., and Elowitz, M.B. (2022). Synthetic mammalian signaling circuits for robust cell population control. Cell *185*, 967-979.e12. https://doi.org/10.1016/j.cell.2022.01.026.

126. Del Vecchio, D., Qian, Y., Murray, R.M., and Sontag, E.D. (2018). Future systems and control research in synthetic biology. Annu. Rev. Control *45*, 5–17. https://doi.org/10.1016/j.arcontrol.2018.04.007.

127. Bodzioch, M., and Belmonte-Beitia, J. (2025). Optimal control of a mathematical model of CAR-T cell therapy for glioblastoma. Discrete Contin. Dyn. Syst. - B *30*, 4255–4275. https://doi.org/10.3934/dcdsb.2025080.

128. Montoya, M., Gallus, M., Phyu, S., Haegelin, J., De Groot, J., and Okada, H. (2024). A Roadmap of CAR-T-Cell Therapy in Glioblastoma: Challenges and Future Perspectives. Cells *13*, 726. https://doi.org/10.3390/cells13090726.

129. Tu, Z., Chen, Y., Zhang, Z., Meng, W., and Li, L. (2025). Barriers and solutions for CAR-T therapy in solid tumors. Cancer Gene Ther. https://doi.org/10.1038/s41417-025-00931-7.

130. Jindal, V., Akella, P., Piranavan, P., and Siddiqui, A.D. (2019). A mixed method analysis of toxicities of anti-CD19-CAR T cell therapy in hematological malignancies. J. Clin. Oncol. *37*, e19068–e19068. https://doi.org/10.1200/JCO.2019.37.15_suppl.e19068.

131. Xiao, X., Huang, S., Chen, S., Wang, Y., Sun, Q., Xu, X., and Li, Y. (2021). Mechanisms of cytokine release syndrome and neurotoxicity of CAR T-cell therapy and associated prevention and management strategies. J. Exp. Clin. Cancer Res. *40*, 367. https://doi.org/10.1186/s13046-021-02148-6.

132. Yan, Z., Zhang, H., Cao, J., Zhang, C., Liu, H., Huang, H., Cheng, H., Qiao, J., Wang, Y., Wang, Y., et al. (2021). Characteristics and Risk Factors of Cytokine Release Syndrome in Chimeric Antigen Receptor T Cell Treatment. Front. Immunol. *12*, 611366. https://doi.org/10.3389/fimmu.2021.611366.

133. Velasco, R., Mussetti, A., Villagrán-García, M., and Sureda, A. (2023). CAR T-cell-associated neurotoxicity in central nervous system hematologic disease: Is it still a concern? Front. Neurol. *14*, 1144414. https://doi.org/10.3389/fneur.2023.1144414.

134. Jones, D.K., Eckhardt, C.A., Sun, H., Tesh, R.A., Malik, P., Quadri, S., Firme, M.S., Van Sleuwen, M., Jain, A., Fan, Z., et al. (2022). EEG-based grading of immune effector cell-associated neurotoxicity syndrome. Sci. Rep. *12*, 20011. https://doi.org/10.1038/s41598-022-24010-1.

135. Majzner, R.G., and Mackall, C.L. (2018). Tumor Antigen Escape from CAR T-cell Therapy. Cancer Discov. *8*, 1219–1226. https://doi.org/10.1158/2159-8290.CD-18-0442.

136. Lin, H., Yang, X., Ye, S., Huang, L., and Mu, W. (2024). Antigen escape in CAR-T cell therapy: Mechanisms and overcoming strategies. Biomed. Pharmacother. *178*, 117252. https://doi.org/10.1016/j.biopha.2024.117252.

137. Rodriguez-Garcia, A., Palazon, A., Noguera-Ortega, E., Powell, D.J., and Guedan, S. (2020). CAR-T Cells Hit the Tumor Microenvironment: Strategies to Overcome Tumor Escape. Front. Immunol. *11*, 1109. https://doi.org/10.3389/fimmu.2020.01109.

138. Chen, T., Wang, M., Chen, Y., and Liu, Y. (2024). Current challenges and therapeutic advances of CAR-T cell therapy for solid tumors. Cancer Cell Int. *24*, 133. https://doi.org/10.1186/s12935-024-03315-3.

139. Cheever, A., Townsend, M., and O'Neill, K. (2022). Tumor Microenvironment Immunosuppression: A Roadblock to CAR T-Cell Advancement in Solid Tumors. Cells *11*, 3626. https://doi.org/10.3390/cells11223626.

140. Safarzadeh Kozani, P., Safarzadeh Kozani, P., and Rahbarizadeh, F. (2022). Addressing the obstacles of CAR T cell migration in solid tumors: wishing a heavy traffic. Crit. Rev. Biotechnol. *42*, 1079–1098. https://doi.org/10.1080/07388551.2021.1988509.

141. Chen, Z., Pan, H., Luo, Y., Yin, T., Zhang, B., Liao, J., Wang, M., Tang, X., Huang, G., Deng, G., et al. (2021). Nanoengineered CAR-T Biohybrids for Solid Tumor Immunotherapy with Microenvironment Photothermal-Remodeling Strategy. Small *17*, 2007494. https://doi.org/10.1002/smll.202007494.

142. Bonifant, C.L., Jackson, H.J., Brentjens, R.J., and Curran, K.J. (2016). Toxicity and management in CAR T-cell therapy. Mol. Ther. - Oncolytics *3*, 16011. https://doi.org/10.1038/mto.2016.11.

143. Tur, C., Eckstein, M., Velden, J., Rauber, S., Bergmann, C., Auth, J., Bucci, L., Corte, G., Hagen, M., Wirsching, A., et al. (2025). CD19-CAR T-cell therapy induces deep tissue depletion of B cells. Ann. Rheum. Dis. *84*, 106–114. https://doi.org/10.1136/ard-2024-226142.

144. Kochenderfer, J.N., Dudley, M.E., Feldman, S.A., Wilson, W.H., Spaner, D.E., Maric, I., Stetler-Stevenson, M., Phan, G.Q., Hughes, M.S., Sherry, R.M., et al. (2012). B-cell depletion and remissions of malignancy along with cytokine-associated toxicity in a clinical trial of anti-CD19 chimeric-antigen-receptor–transduced T cells. Blood *119*, 2709–2720. https://doi.org/10.1182/blood-2011-10-384388.

145. Gumber, D., and Wang, L.D. (2022). Improving CAR-T immunotherapy: Overcoming the challenges of T cell exhaustion. eBioMedicine *77*, 103941. https://doi.org/10.1016/j.ebiom.2022.103941.

146. Kim, S.J., Yoon, S.E., and Kim, W.S. (2024). Current Challenges in Chimeric Antigen Receptor T-cell Therapy in Patients With B-cell Lymphoid Malignancies. Ann. Lab. Med. *44*, 210–221. https://doi.org/10.3343/alm.2023.0388.

147. López-Cantillo, G., Urueña, C., Camacho, B.A., and Ramírez-Segura, C. (2022). CAR-T Cell Performance: How to Improve Their Persistence? Front. Immunol. *13*, 878209. https://doi.org/10.3389/fimmu.2022.878209.

148. Xiong, D., Yu, H., and Sun, Z.-J. (2024). Unlocking T cell exhaustion: Insights and implications for CAR-T cell therapy. Acta Pharm. Sin. B *14*, 3416–3431. https://doi.org/10.1016/j.apsb.2024.04.022.

149. Dreyzin, A., Rankin, A.W., Luciani, K., Gavrilova, T., and Shah, N.N. (2024). Overcoming the challenges of primary resistance and relapse after CAR-T cell therapy. Expert Rev. Clin. Immunol. *20*, 745–763. https://doi.org/10.1080/1744666X.2024.2349738.

150. Maus, M.V., and June, C.H. (2016). Making Better Chimeric Antigen Receptors for Adoptive T-cell Therapy. Clin. Cancer Res. *22*, 1875–1884. https://doi.org/10.1158/1078-0432.CCR-15-1433.

151. Wherry, E.J., and Kurachi, M. (2015). Molecular and cellular insights into T cell exhaustion. Nat. Rev. Immunol. *15*, 486–499. https://doi.org/10.1038/nri3862.

152. Pauken, K.E., and Wherry, E.J. (2015). Overcoming T cell exhaustion in infection and cancer. Trends Immunol. *36*, 265–276. https://doi.org/10.1016/j.it.2015.02.008.

153. Long, A.H., Haso, W.M., Shern, J.F., Wanhainen, K.M., Murgai, M., Ingaramo, M., Smith, J.P., Walker, A.J., Kohler, M.E., Venkateshwara, V.R., et al. (2015). 4-1BB costimulation ameliorates T cell exhaustion induced by tonic signaling of chimeric antigen receptors. Nat. Med. *21*, 581–590. https://doi.org/10.1038/nm.3838.

154. Kaech, S.M., and Cui, W. (2012). Transcriptional control of effector and memory CD8+ T cell differentiation. Nat. Rev. Immunol. *12*, 749–761. https://doi.org/10.1038/nri3307.

155. Pan, K., Farrukh, H., Chittepu, V.C.S.R., Xu, H., Pan, C., and Zhu, Z. (2022). CAR race to cancer immunotherapy: from CAR T, CAR NK to CAR macrophage therapy. J. Exp. Clin. Cancer Res. *41*, 119. https://doi.org/10.1186/s13046-022-02327-z.

156. Rivas, C.H., Cole, A., Rooney, C.M., and Parihar, R. (2018). Abstract 4734: Suppressive myeloid cells of the solid tumor microenvironment enhance regulatory T cell function and differentially affect CAR-T cell function. Cancer Res. *78*, 4734–4734. https://doi.org/10.1158/1538-7445.AM2018-4734.

157. Xia, X., Yang, Z., Lu, Q., Liu, Z., Wang, L., Du, J., Li, Y., Yang, D.-H., and Wu, S. (2024). Reshaping the tumor immune microenvironment to improve CAR-T cell-based cancer immunotherapy. Mol. Cancer *23*, 175. https://doi.org/10.1186/s12943-024-02079-8.

158. Wang, M., Zhang, C., and Jiang, X. (2021). CAR-T: a potential gene carrier targeting solid tumor immune microenvironment. Signal Transduct. Target. Ther. *6*, 393. https://doi.org/10.1038/s41392-021-00812-z.

159. Rojas-Quintero, J., Díaz, M.P., Palmar, J., Galan-Freyle, N.J., Morillo, V., Escalona, D., González-Torres, H.J., Torres, W., Navarro-Quiroz, E., Rivera-Porras, D., et al. (2024). Car T Cells in Solid Tumors: Overcoming Obstacles. Int. J. Mol. Sci. *25*, 4170. https://doi.org/10.3390/ijms25084170.

160. Chakrabarti, A., and Ghosh, J.K. (2011). AIC, BIC and Recent Advances in Model Selection. In Philosophy of Statistics Handbook of the Philosophy of Science., P. S. Bandyopadhyay and M. R. Forster, eds. (North-Holland), pp. 583–605. https://doi.org/10.1016/B978-0-444-51862-0.50018-6.

161. Brunton, S.L., Proctor, J.L., and Kutz, J.N. (2016). Discovering governing equations from data by sparse identification of nonlinear dynamical systems. Proc. Natl. Acad. Sci. *113*, 3932–3937. https://doi.org/10.1073/pnas.1517384113.

162. Brummer, A.B., Xella, A., Woodall, R., Adhikarla, V., Cho, H., Gutova, M., Brown, C.E., and Rockne, R.C. (2023). Data driven model discovery and interpretation for CAR T-cell killing using sparse identification and latent variables. Front. Immunol. *14*, 1115536. https://doi.org/10.3389/fimmu.2023.1115536.